\documentclass[12pt]{article}

\usepackage{a4wide}
\usepackage{cite}
\usepackage{amssymb}
\usepackage{amsmath}
\usepackage{epsfig}
\usepackage{graphicx}
\usepackage{eucal}

\newlength{\horizMargin}
\newcommand{\asb}{\bar{\alpha}_\mathrm{s}}       
\newcommand{\alh}{\hat{\alpha}}          
\newcommand{\as}{\alpha_\mathrm{s}}      
\newcommand{\BFKL}{\mathrm{BFKL}}        
\newcommand{\G}{\mathcal{G}}             
\newcommand{\chireg}{\chi_{1,\mathrm{reg}}}
\newcommand{\chiht}{\chi_{\mathrm{ht}}}  
\newcommand{\column}[2]{\begin{pmatrix} #1 \\ #2 \end{pmatrix}}
\newcommand{\DIS}{\mathrm{DIS}}          
\newcommand{\dif}{\mathrm{d}}            
\newcommand{\Dqg}{\Delta_{qg}}

\newcommand{\dqg}{\delta_{qg}}

\newcommand{\eff}{\mathrm{eff}}          
\newcommand{\Ga}{\Gamma}
\newcommand{\GeV}{\,\text{GeV}}
\newcommand{\ga}{\gamma}
\newcommand{\gas}{\gamma_{\mathrm{s}}}   

\newcommand{\half}{{\textstyle\frac{1}{2}}}
\newcommand{\cK}{\mathsf{K}}
\newcommand{\Kht}{K_{\mathrm{ht}}}       
\newcommand{\kt}{\boldsymbol{k}}         
\newcommand{\Li}{{\mathrm{Li}_2}}        
\newcommand{\LL}{{\mathrm{LLx}}}
\newcommand{\MSbar}{\overline{\mathrm{MS}}}
\newcommand{\NL}{{\mathrm{NL}x}}
\newcommand{\om}{\omega}
\newcommand{\omhalf}{{\textstyle\frac{\omega}{2}}}
\newcommand{\oms}{\omega_{\mathrm{s}}}   
\newcommand{\NC}{N_C}
\newcommand{\CF}{C_F}
\newcommand{\CA}{C_A}
\newcommand{\Tf}{T_f}
\newcommand{\nf}{n_f}
\newcommand{\Tr}{T_r}
\DeclareMathOperator{\tr}{\mathrm{tr}}   
\newcommand{\qt}{\boldsymbol{q}}         
\newcommand{\row}[2]{\begin{pmatrix} #1 & #2 \end{pmatrix}}
\newcommand{\run}{\mathrm{run}}
\newcommand{\ugd}{{\CMcal F}}            
\newcommand{\ui}{\mathrm{i}}             
\newcommand{\order}[1]{{\cal O}\left(#1\right)}

\newcommand{\zero}[1]{{}_{0}#1}
\newcommand{\one}[1]{{}_{1}#1}
\newcommand{\two}[1]{{}_{2}#1}

\newcommand{\npb}[3]{{\it Nucl.~Phys.~}{\bf B #1} (#2) #3}

\newcommand{\plb}[3]{{\it Phys.~Lett.~}{\bf B #1} (#2) #3}
\newcommand{\prd}[3]{{\it Phys.~Rev.~}{\bf D #1} (#2) #3}

\newcommand{\zpc}[3]{{\it Z.~Phys.~}{\bf C #1} (#2) #3}
\numberwithin{equation}{section}

\begin{document}


\titlepage

\begin{flushright} CERN-PH-TH/2007-104\\
  DFF 436/07/07\\ 
  July 2007
\end{flushright}

\vspace*{1in}

\begin{center}
  {\Large \bf
  A matrix formulation for small-$\boldsymbol{x}$ singlet evolution}
  
  \vspace*{0.4in}
  
  M.~Ciafaloni$^{(a)}$\footnote{On sabbatical leave of absence from
  Dipartimento di 
  Fisica, Universit\`a di Firenze and INFN, Sezione di Firenze.},
  D.~Colferai$^{(b)}$,
  G.P.~Salam$^{(c)}$
  and A.M.~Sta\'sto$^{(d)}$ \\
  
  {\small
  \vspace*{0.5cm}
  $^{(a)}$ {\it CERN, Department PH-TH, CH-1211 Geneva 23, Switzerland}\\
  \vskip 2mm
  $^{(b)}$ {\it  Dipartimento di Fisica, Universit\`a di Firenze,
   50019 Sesto Fiorentino (FI), Italy}; \\
  {\it  INFN Sezione di Firenze,  50019 Sesto Fiorentino (FI), Italy}\\
  \vskip 2mm
  $^{(c)}$ {\it LPTHE, Universit\'e P. et M. Curie -- Paris 6,
  Universit\'e D. Diderot -- Paris 7, CNRS~UMR~7589,
  Paris, France}\\
  \vskip 2mm
  $^{(d)}$ {\it Department of Physics, Pennsylvania State University, University Park, 16802 PA, USA};\\
  \vskip 2mm
  {\it H.~Niewodnicza\'nski Institute of Nuclear Physics, Krak\'ow, Poland}\\
  \vskip 2mm}
\end{center}


\bigskip

\begin{abstract}
  \noindent
  We propose a matrix evolution equation in $(x,\kt)$-space for flavour
  singlet, unintegrated quark and gluon densities, which generalizes DGLAP and
  BFKL equations in the relevant limits. The matrix evolution kernel is
  constructed so as to satisfy renormalization group constraints in both the
  ordered and antiordered regions of exchanged momenta $\kt$, and
  incorporates the known NLO anomalous dimensions in the $\MSbar$ scheme as
  well as the NL$x$ BFKL kernel. We provide a hard Pomeron exponent and
  effective eigenvalue functions that include the $n_f$-dependence,
  and give also the
  matrix of resummed DGLAP splitting functions. The results connect smoothly
  with those of the single-channel approach. The novel $P_{qa}$
  splitting functions show resummation effects delayed down to $x=10^{-4}$,
  while both $P_{ga}$ entries show a shallow dip around $x=10^{-3}$, similarly
  to the $gg$ single-channel results. We remark that the matrix formulation
  poses further constraints on the consistency of a BFKL framework with the
  $\MSbar$ scheme, which are satisfied at NLO, but marginally violated by
  small $n_f/N_c^2$-suppressed terms at NNLO.
\end{abstract}

\newpage

\section{Introduction\label{sec:intro}}

Small-x QCD evolution, historically based on DGLAP~\cite{DGLAP} and BFKL~\cite{BFKL} dynamics,
has been widely investigated in the past years~\cite{SmallxColl,RGvert,QQvert,FaLi98,CaCi98,CaHa94,Salam98}, leading to a better understanding
of the two approaches just mentioned and to robust resummed predictions for the gluon density
and splitting function~\cite{ omExp,rgiggf,extending,gluonsf,CCSS_MSbar,ABF_fit,ABF_improved,THORNE}.
There is now a remarkable consensus~\cite{hera_lhc_ws} among various resummation
approaches on the resulting gluon evolution kernel, and a satisfactory
comparison of some of them to experimental data~\cite{ABF_fit,ThWh06}.

The basic idea underlying the progress of the resummation approaches just
mentioned, lies in the observation~\cite{omExp} that the BFKL kernel embodies an
infinite number of subleading contributions which are collinear singular.
These terms are parametrically large and need  to be taken into account in
order to achieve consistency with the renormalization group (RG). The
techniques for incorporating such terms differ in detail according to the
various authors, but lead eventually to similar results.

However, all approaches developed so far limit themselves to a consistent
resummation scheme only for the evolution of the gluon density.  The quark-sea
contribution is obtained on the basis of $\kt$-factorization of the $ q\bar q$
dipole~\cite{QQvert} in the DIS factorization scheme and/or by adding the
$q\bar q$ contribution to the next-to-leading-$\log x$ (NL$x$) BFKL
kernel. The main purpose of the present paper is to devise a resummed
small-$x$ evolution scheme in a coupled matrix form, so as to treat gluons and
quarks on the same footing and in a collinear factorization scheme which is as
close as possible to a predetermined one, e.g.\ the $\MSbar$ scheme.

Our matrix approach, in the collinear limit, has the advantage that it
complies automatically with the matrix factorization of the integrated
partonic densities in the singlet evolution, and thus is able to incorporate
the known low-order anomalous dimensions for any value of $\om=N-1$, the
moment index. On the other hand, in the high-energy limit, the
(gauge-invariant) unintegrated partonic densities are well defined by
$\kt$-factorization around different values of $\om$. To be precise, the gluon
unintegrated density is defined around $\om=0$ and the quark around
$\om=-1$. Therefore, in the leading high energy region --- that is around
$\om=0$ --- we are able to include the known LL$x$+NL$x$ BFKL kernel in the
gluon channel {\em only}, thus leaving the quark entries somewhat
unconstrained from the $\kt$-factorization standpoint.

Note however that assuming a BFKL framework in matrix form is a demanding
requirement, because $\kt$-factorization implies resummation formulae for the
anomalous dimension matrix up to NL$x$ level. Therefore, incorporating both
exact low-order anomalous dimensions (say, in the $\MSbar$ scheme) and NL$x$
expressions (from the exact BFKL kernel) imposes on our matrix kernel some
nontrivial consistency relations expressing the requirement that collinear and
high-energy schemes do not conflict with each other. They are discussed in detail
in the following, and we find that they are satisfied in the $\MSbar$ scheme
up to NLO level, while they are marginally violated at NNLO, by small terms in
the $gq$ entry of relative order $n_f/N_c^2$. For this reason we restrict
ourselves, in this paper, to the NLO-NL$x$ level. This means that, starting at
NNLO, our splitting functions will be in some matrix scheme which is not the
$\MSbar$ scheme. Higher order effects on the scheme change have been studied
in~\cite{CCSS_MSbar} and have been found to be comparable to renormalization scale
uncertainties.

Besides the NLO-NL$x$ information mentioned above, we impose the general
requirement of consistency with the renormalization group in both ordered and
antiordered configurations of exchanged partonic momenta $\kt$'s. This is best
expressed in the $\ga \leftrightarrow 1+\om-\ga$ symmetry of the
kernel, where $\ga$ is conjugated to $\log\kt^2$. We enforce this symmetry by
the so-called consistency constraint~\cite{AGS,KMS,Ciaf88} which introduces an
$\om$-dependence in the (leading) kernel, so as to resum those
parts~\cite{Salam98} of the higher order BFKL kernels which are required by the
RG. This procedure follows previous papers~\cite{omExp,rgiggf} and is used in
particular for the $gg$ matrix element of the kernel.

Despite all such requirements, there is a considerable ambiguity in our approach
which is tied up to the matrix structure, because the $\ga$ and $\om$
dependences of the various matrix elements of the kernel are constrained only to
a limited extent by the collinear and high-energy limits.  We
therefore introduce in 
sec.~\ref{sec:bmf} some further requirements, mostly related to the pole
structure of the kernel in the $\ga$ and $\om$ variables, by requiring it
to have at most simple poles. This assumption is quite natural in the case of
$\om \rightarrow 0$ because of the BFKL limit, and follows by the
$\om$-expansion method~\cite{omExp} in the $\ga\rightarrow 0$ case. The
leading twist pole structure of the kernel at lowest order in $\as$ is basically
$\cK \sim \Ga_0\; \left[(1/\ga) + 1/(1+\om-\ga) \right]$, where
$\Ga_0$ denotes the LO DGLAP anomalous dimension matrix. The full kernel to
second order in $\as$ is constructed in secs.~\ref{sec:bmf} and \ref{sec:knla}
according to the requirements stated above. It also contains running coupling
effects at the scales suggested by the NL$x$ BKL kernel and by the RG.

In the frozen $\as$ case we calculate in sec.~\ref{s:fcad} the resulting
anomalous dimension matrix, and its eigenvalues $\ga_{\pm}(\as,\om)$. The
leading eigenvalue at high energies, $\ga_+$, contains important resummation
effects in the $\as/\om$ variable. We also obtain resummation formulae for
$\Ga_{qg}$ and $\Ga_{gq}$, the latter being specific to our matrix approach and
not directly obtained on the basis of the NL$x$ BFKL kernel only. In
sec.~\ref{s:cfrgf} we present results for the effective eigenvalue (or
characteristic) functions $\om=\chi_{\pm}(\as,\ga)$ as inverse functions of the
anomalous dimensions,
and for the hard Pomeron exponent $\oms(\as)$.
In the case with running coupling described in sec.~\ref{s:nrrc}, we provide the
resummed DGLAP splitting function matrix in $x$ space, obtained by the
numerical deconvolution method proposed in~\cite{CCSFact,rgiggf} and generalized to
the matrix case. Details of the matrix kernel and of the anomalous dimension
expressions are left to Appendices A-C.

\section{Basis of matrix formulation\label{sec:bmf}}

The general purpose of this paper is to provide integro-differential
matrix equations for unintegrated parton distributions, which
interpolate between DGLAP evolution equations~\cite{DGLAP} in the hard
scale variable $\log \kt^2$ and the high-energy BFKL evolution
equation~\cite{BFKL} in the rapidity-like variable $\log 1/x$.
Despite the high-energy and collinear factorization constraints, the
above interpolation is subject to considerable ambiguities, due to the
following facts:
\begin{itemize}
\item[(a)] Off-shell, unintegrated densities are defined in a gauge-invariant
  way by
  $\kt$-factorization of gluon and quark exchanges around different values of
  the moment index $\om \equiv N-1$, namely $\om = 0$ for the gluon
  and $\om = -1$ for the quark.%
  \footnote{This is because of the spin $1/2$ of the quark exchange, which
    leads to an energy dependence of the cross-section of type $s^{-1}$ at
    high energies. For alternative approaches to the definition of
    unintegrated densities see \cite{Ciaf88,unintegrated}.}  
    Therefore, in the
    high-energy region --- that is around the leading value $\om = 0$ --- only
    some effective gluon equation (which incorporates the high-energy quark
    contributions) is constrained by the BFKL limit, whose kernel has been
    calculated perturbatively~\cite{BFKL,RGvert,QQvert,FaLi98,CaCi98}.  This
    makes the interpolation of the kernel to generic $\om$ values more
    ambiguous for the quark entries.
\item[(b)] The collinear limit constrains the matrix kernel for both
  quarks and gluons, but in a factorization-scheme dependent way, and
  only in the strongly ordered region of transverse momenta
  $\cdots \gg \kt_1^2 \gg \kt_2^2 \gg \cdots$ and in the anti-ordered
  one.  This limit only restricts the singularities of the kernel in
  the variable $\ga$ (conjugated to $\log \kt^2$) so as to reproduce
  the low order anomalous dimension matrix. In addition, the collinear
  $\leftrightarrow$ anti-collinear relationship implies the existence of a
  $\ga \leftrightarrow 1+\om-\ga$ symmetry, whose form is however
  quite general, depending again on the factorization-scheme.
\end{itemize}

\subsection{Basic criteria for the kernel construction\label{ss:bckc}}

In order to tame the ambiguities of the off-shell continuation
mentioned above, we shall use a few basic criteria which --- we shall
argue --- can be consistently imposed and correspond to a
factorization-scheme choice for both high-energy and collinear limits.

Let us refer to a matrix kernel $\cK_{ab}(\as,\om)$, acting on
$\kt$-space, such that the parton Green's function is given by
\begin{equation}
  \label{eq:greenFunc}
  \G_{ab}(\om;\kt,\kt_0) = \left[ 1-\cK(\as,\om)\right]^{-1}_{ab}(\kt,\kt_0) \;,
  \quad (a,b = q,g) \;.
\end{equation}
In the frozen $\as$ limit, the kernel matrix elements are diagonalised in
$\ga$-space and given by the eigenvalue function $\cK_{ab}(\as,\om,\ga)$. Our
first basic assumption is that in the collinear limit $\ga \to 0$ and $\om$
fixed, the matrix kernel $\cK$ shows simple poles only, in the form of a
$\ga$-expansion
\begin{equation}
  \label{eq:gammaStruct}
  \cK = \frac1{\ga} \, \cK^{(0)}(\as,\om) + \cK^{(1)}(\as,\om)
  + \ga \, \cK^{(2)}(\as,\om) + \order{\ga^2} \;.
\end{equation}
Here, the $1/\ga$ singularity is natural because of the DGLAP limit,
and would be the only term present in a pure evolution equation in
$\log\kt^2$. In fact, it implies (sec.~\ref{ss:fklla}) that the
one-loop anomalous dimension matrix $\Ga_0(\om)$ is given by
$\cK^{(0)}_0(\om)$, the coefficient of the $1/\ga$ pole of lowest
order in $\as$ (while the higher order terms $\Ga_n$ involve
$\cK^{(1)},\;\cK^{(2)},\cdots$ as well). 

Note however that higher powers of $\as/\ga$ could have been present also,%
\footnote{For instance, the rough LL$x$ anomalous dimension relation
  $\ga\simeq\asb/\om$ could 
  be replaced in the subleading $\om$-dependence, thus producing higher powers
  of $\as/\ga$.}
and do actually occur in the normal formulation of the NL$x$ BFKL
kernel~\cite{FaLi98,CaCi98}.  By eq.~(\ref{eq:gammaStruct}) we explicitly
exclude such possibility in our matrix kernel, while the BFKL kernel will be
recovered by proper algebraic manipulations (sec.~\ref{sec:knla}).

Our second assumption is analogous to~(\ref{eq:gammaStruct}) with $\om$ and
$\ga$ interchanged. In the high-energy limit of $\om \to 0$ with $\ga$ kept
fixed we require simple pole singularities in the $\om$-expansion
\begin{equation}
  \label{eq:omegaStruct}
  \cK = \frac1{\om} \; \zero\cK(\as,\ga) + \one\cK(\as,\ga)
  + \om \;\two{\cK}(\as,\ga) + \order{\om^2} \;,
\end{equation}
where, in addition,
\begin{equation}
  \label{eq:quarkRowCond}
  \zero\cK_{qq} = 0 = \zero\cK_{qg} \;.
\end{equation}
The $1/\om$ singularity is natural because of the BFKL limit and would
be the only term present in a pure evolution equation in $\log 1/x$.
It implies that the eigenvalue function of the LL$x$ BFKL kernel is given by
$\chi_0(\ga) \sim \zero\cK_0(\ga)$, the coefficient of the $1/\om$ singularity
of lowest order in $\as$ (while the NL$x$ BFKL kernel, discussed in
sec.~\ref{sec:knla}, involves $\one\cK(\ga)$ also).

Therefore, higher order singularities in $\om$ --- which are present
in the anomalous dimension at higher order --- will be obtained
(sec.~\ref{s:fcad}) by using the rough anomalous dimension relation
$\ga \simeq \asb/\om$ in the subleading $\ga$-dependence. The fact that
only $\zero\cK_{gq}$ and $\zero\cK_{gg}$ possess the $1/\om$
singularity is related to the fact that in usual factorization schemes~\cite{CaHa94}, only
$\Ga_{gq}$ and $\Ga_{gg}$ show a LL$x$ dependence on the $\as/\om$
variable.

There is a third important assumption, which deals with the
relationship between collinear and anti-collinear orderings of
exchanged transverse momenta. Both orderings are to be incorporated in
our off-shell formulation and simple kinematical considerations show
that the variable conjugated to $\log \kt^2$ in the reverse ordering
is $1+\om-\ga$. Therefore, values of $\ga$ and $1+\om-\ga$ must be
related by some symmetry, and we shall assume, out of simplicity,
\begin{equation}
  \label{eq:symmetry}
  \cK_{ab}(\ga,\om) = \cK_{ab}(1+\om-\ga,\om) \;.
\end{equation}

It is perhaps useful to recall the formal basis for the
symmetry~(\ref{eq:symmetry}). Let us write the $\kt$-factorization formula for
the $A+B \to X$ differential cross-section in the form~\cite{CiaCo98}
\begin{equation}
  \label{heFact}
  \frac{\dif\sigma^{AB}}{\dif^2\kt \, \dif^2\kt_0} = \int \frac{\dif\om}{2\pi\ui}
  \left(\frac{s}{k k_0}\right)^\om h^A(\kt,\om) \, \G(\kt,\kt_0;\om) \,
  h^B(\kt_0,\om) \;, \quad  (k \equiv |\kt|) \;,
\end{equation}
where we have lumped in the $A, B$ superscripts the dependence on the hard
scales $Q_A, Q_B$ of the process.
Then the change of energy-scale from $kk_0$ to, say, $k^2$ can be incorporated
by the change of kernel
\begin{equation}
  \label{kernelChange}
  \cK^{[k^2]}(\kt,\kt';\om)
  = \left(\frac{k}{k'}\right)^\om \cK^{[k k_0]}(\kt,\kt';\om) \; ,
\end{equation}
or, at frozen $\as$, by the $\om$-dependent shift~\cite{CaCi98, Salam98} of the
corresponding eigenvalue functions
\footnote{We use the symbol $\chi$ to denote eigenvalue functions of kernels
  considered in or related to previous works on small-$x$ resummations in the
  gluon-channel. We keep the symbol $\cK$ to denote both kernels and eigenvalue
  functions specifically designed for this matrix formulation. Note also that
  the $\chi$'s are perturbative coefficients of expansions in
  $\asb\equiv \as\CA/\pi$, therefore differing in normalization by a factor
  $(2\CA)^{-(n+1)}$ from their $\cK$ counterparts in
  eq.~(\ref{eq:Kexpansion}), because $\alh=\as/2\pi$.}
\begin{equation}
  \label{eigenvChange}
  \chi^{[k^2]}(\ga,\om) = \chi^{[k k_0]}(\ga-\omhalf,\om) \, .
\end{equation}
On the other hand, in the one-channel case --- namely when one considers only
gluon dynamics --- the $A \leftrightarrow B$ symmetry of $\sigma_{AB}$ implies
the $\kt \leftrightarrow \kt'$ symmetry of the kernel $\cK(\kt,\kt';\om)$ and the
$\ga \leftrightarrow 1-\ga$ symmetry of the eigenvalue functions
$\chi^{[k k_0]}(\ga,\om)$. Therefore, at energy-scale $k^2$,
the $\ga \leftrightarrow 1+\om-\ga$ symmetry of eq.~(\ref{eq:symmetry}) holds
for $\chi^{[k^2]}(\ga,\om)$, whose superscript will be dropped from now on.

In the matrix case, the thorough discussion of
sec.~\ref{ss:gfdrs} shows that the collinear $\leftrightarrow$ anti-collinear symmetry of the
matrix kernel is expected to have the more general form
\begin{equation}
  \label{eq:genSymmetry}
  \cK(1+\om-\ga,\om) = S(\om) \cK^{T}(\ga,\om) S^{-1}(\om) \;.
\end{equation}
Therefore, eq.~(\ref{eq:symmetry}) is obtained by choosing the similarity
transformation $S$ so as to have
\begin{equation}
\label{eq:similarity}
 S \cK^{T} S^{-1} = \cK \; ,
\end{equation}
and represents yet another restriction of our off-shell scheme.%
\footnote{The choice of eq.~(\ref{eq:symmetry}) must be supplemented by a
  corresponding choice of impact factors in order to satisfy the
  symmetry for observable cross-sections. In the realistic NLO-NL$x$ case the
  similarity transformation $S$ in eq.~(\ref{eq:similarity}) is expected to be
  an operator in $\kt$-space also.}  

In the following we shall show in more detail how to construct the
matrix kernel so as to satisfy the known collinear/high-energy limits
with LO-LL$x$ accuracy (secs.~\ref{ss:fklla}, \ref{ss:gfdrs}) and
NLO-NL$x$ accuracy (sec.~\ref{sec:knla}), within the scheme
restrictions provided by
assumptions~(\ref{eq:gammaStruct}, \ref{eq:omegaStruct}, \ref{eq:symmetry}).
We note from the start that eqs.~(\ref{eq:gammaStruct}) and
(\ref{eq:omegaStruct}) impose  consistency relations on the
anomalous dimensions, which show up in a novel NL$x$ resummation
formula for $\Ga_{gq}$, to be discussed in detail in
secs.~\ref{ss:rf} and \ref{ss:cr}.

\subsection{Form of kernel at LO-LL$\boldsymbol x$ accuracy\label{ss:fklla}}

In order to discuss the above features in more detail in the frozen
$\as$ limit, we introduce the triple expansion
\begin{equation}
  \label{eq:Kexpansion}
  \cK(\as,\ga,\om) \equiv \sum_{n,m,p=0}^{\infty} {}_p\cK_n^{(m)} \;
  \alh^{n+1} \ga^{m-1} \om^{p-1} \;, \qquad \alh \equiv \frac{\as}{2\pi} \;,
\end{equation}
where $_p\cK_n^{(m)}$ are $2 \times 2$ matrices in the $a=q,g$ indices, and we
note that $m,p \ge 0$, consistently with the simple pole assumption of
eqs.~(\ref{eq:gammaStruct}, \ref{eq:omegaStruct}). We also use the notation
$\cK^{(m)}(\as,\om)$, $_p\cK(\as,\ga)$, $\cK_n^{(m)}(\om)$
and $_p\cK_n(\ga)$ to mean partially resummed
coefficients, as already done before. In this paper we limit ourselves to two
terms ($n=0,1$) in the frozen $\as$-expansion, which will be able to
accommodate the LL$x$ and NL$x$ BFKL kernels. However, running coupling
effects will be introduced by various scale choices for the various terms
(sec.~\ref{ss:runningc}) and this implies in general an infinite series when
expanding around a fixed scale.

Let us first show how to construct $\cK_0$ so as to be consistent with
the collinear and high-energy limit at LO-LL$x$ accuracy. We denote
by $\Ga$ the anomalous dimension matrix, with the expansion
\begin{equation}
  \label{eq:GammaExpansion}
  \Ga(\om) \equiv \sum_{n=0}^\infty \alh^{n+1} \Ga_n(\om) \;,
\end{equation}
where we recall the small-$\om$ behaviour ($\Tf \equiv \Tr \nf=\nf/2$)
\begin{equation}
  \label{eq:Gamma0smallx}
  \Ga_0(\om) =
  \begin{pmatrix}
    \order{\om} & \displaystyle{\frac{4\Tf}{3}} + \order{\om} \\ & \\
    \displaystyle{\frac{2\CF}{\om}} + \order{1} & \displaystyle{\frac{2\CA}{\om}} + \order{1}
  \end{pmatrix}\;,
\end{equation}
and, in the $\MSbar$ scheme,
\begin{equation}
  \label{eq:Gamma1smallx}
  \Ga_1(\om) = \frac1{9\om}
  \begin{pmatrix}
    40 \Tf \CF & 40 \Tf \CA \\ & \\
    9 \CF \CA - 40 \Tf \CF & (12 \CF - 46 \CA) \Tf
  \end{pmatrix}\;,
\end{equation}
with the two eigenvalues
\begin{align}
  \ga_{+,0} &= \frac{2\CA}{\om} + \order{1} \;, \quad
  \ga_{+,1} = -\frac{2\Tf}{9\om} \left( 10 \CA - \frac{13}{\CA} \right) + \order{1}
  \label{eq:plusEigenv} \\
  \ga_{-,0} &= -\frac{4 \Tf \CF}{3 \CA} + \order{\om} \;.
\end{align}

We then write the generalised BFKL equation for the unintegrated
parton densities $\ugd_i(\kt;\om)$ in the form
\begin{equation}
  \label{eq:genBFKLeq}
  \ugd = \cK \ugd + \ugd^{\mathrm{source}} \;,
\end{equation}
where the source $\ugd^{\mathrm{source}}$ is local in $\kt$-space. It is then
straightforward (cf.~sec.~\ref{ss:adm} and app.~\ref{a:read}) to derive, for
frozen $\as$, DGLAP type equations for the integrated densities
\begin{equation}
  \label{eq:integratedDensities}
  f_i(Q^2;\om) \equiv \int^{Q^2} \dif^2 \kt \; \ugd_i(\kt;\om) \; ,
  \qquad (i = q,g)
\end{equation}
of type
\begin{equation}
  \label{eq:DGLAPeq}
  \dot{f}_i \equiv \frac{\partial f_i}{\partial\log Q^2}
  = \sum_{j=q,g} \Ga_{ij} f_j \;,
\end{equation}
where
\begin{subequations}\label{eq:Gamma_from_cK}
\begin{align}
  \Ga_0 &= \cK_0^{(0)}
 \label{Gamma_0} \\
  \Ga_1 &= \cK_1^{(0)} + \cK_0^{(1)}\cK_0^{(0)}
 \label{Gamma_1} \\
  \Ga_2 &= \cK_2^{(0)} + \cK_1^{(1)}\cK_0^{(0)} + \cK_0^{(1)}\cK_1^{(0)}
            + \cK_0^{(2)} \big(\cK_0^{(0)}\big)^2 + \big(\cK_0^{(1)}\big)^2\cK_0^{(0)} \; ,
 \label{Gamma_2}
\end{align}
\end{subequations}
and so on. This identification relies on the
expansion~(\ref{eq:Kexpansion}), which in turn relies on the assumed
single $\ga$-pole structure of~(\ref{eq:gammaStruct}), see app.~\ref{a:read}.

We note that this procedure implies that the  part of the anomalous dimensions,  proportional to $\cK^{(i)}$ with $i>0$, at a given order of perturbation theory  can be generated from the lower orders.

Eqs.~(\ref{eq:Gamma_from_cK}) can be used to constrain recursively the
$\ga \to 0$ singularities of $\cK_0,\;\cK_1,\cdots$ for a given set of low
order anomalous dimensions, for instance in the $\MSbar$ scheme. In
particular, it determines $\cK_0^{(0)} = \Ga_0$, as noticed after
eq.~(\ref{eq:gammaStruct}), but does not fix $\cK_0^{(1)}$, which is
therefore a scheme-changing parameter. At LO-LL$x$ level, we choose
the parameterization
\begin{equation}
  \label{eq:K0param}
  \cK_0(\ga,\om) =
  \begin{pmatrix}
    \Ga_{qq,0}(\om) \chi_c^\om(\ga) & \Ga_{qg,0}(\om) \chi_c^\om(\ga) \\ & \\
    \Ga_{gq,0}(\om) \chi_c^\om(\ga) &
    \frac{2\CA}{\om} \chi_0^\om(\ga)
    + \left[\Ga_{gg,0}(\om)-\frac{2\CA}{\om}\right]\chi_c^\om(\ga) 
  \end{pmatrix} \; ,
\end{equation}
where
\footnote{We will use the $\om$ superscript to denote
  $\om$-shifted~\cite{rgiggf} eigenvalue functions
  $\chi^\om(\ga)\equiv\chi_L(\ga)+\chi_L(1+\om-\ga)$, $\chi_L$ being
  the ``left projection'' of the eigenvalue function
  $\chi(\ga) = \chi_L(\ga)+\chi_L(1-\ga)$ with
  singularities in the half-plane $\Re(\ga) < 1/2$ only.}
\begin{equation}
  \label{eq:chi_0}
  \chi_0^\om(\ga) \equiv 2\psi(1) - \psi(\ga) - \psi(1+\om-\ga) \; ,
\end{equation}
reduces to the leading BFKL eigenvalue function in the $\om = 0$
limit ($\psi(x)$ is the Digamma function), and
\begin{equation}
  \label{eq:chi_c}
    \chi_c^\om(\ga) = \frac{1}{\ga} + \frac{1}{1+\om-\ga} \; ,
\end{equation}
is a simple interpolation of $\ga = 0$ and $1+\om-\ga = 0$ poles. Higher twist
terms will be needed at the NL level in  sec.~\ref{sec:knla}.

Let us note that in eq.~(\ref{eq:K0param}) we have already
incorporated assumptions~(\ref{eq:omegaStruct}) and
(\ref{eq:symmetry}) at this level. In fact, the form~(\ref{eq:K0param})
is consistent with eq.~(\ref{eq:omegaStruct}) because of the BFKL limit
of $\chi_0^\om$ in eq.~(\ref{eq:chi_0}) and of the absence of $1/\om$
poles in $\cK_{qq}$ and $\cK_{qg}$. Furthermore, the
symmetry~(\ref{eq:symmetry}) is present by construction. On the other
hand, the simple form of $\chi_c^\om$ in eq.~(\ref{eq:chi_c}) and the
fact that it appears unchanged in the $qq$, $qg$ and $gq$ entries and
in the part of the $gg$ entry which has no $1/\om$ singularities are
all ``off-shell'' features which we choose for simplicity reasons.

The fact that $\cK_{qq}$ and $\cK_{gq}$ are proportional to $\chi_c^\om$ helps
to satisfy the momentum conservation sum rule. Indeed, since
\begin{equation}\label{momConSumRule}
 \Ga_{qq,0}(\om=1)+\Ga_{gq,0}(\om=1)=\Ga_{qg,0}(\om=1)+\Ga_{gg,0}(\om=1)=0 \;,
\end{equation}
we obtain
\begin{equation}
  \label{eq:QplusGdot}
  \dot{f}_q + \dot{f}_g \equiv \ugd_q + \ugd_g = \chi_{ht} \, \ugd_g \;,
\end{equation}
where all quantities are evaluated at $\om = 1$ and $\chi_{ht}$ is
some higher-twist kernel, having singularities at
$\ga = -1, -2, \cdots$. It is then easy to show that the sum-rule
violation is at most $\order{\as^2}$, instead of $\order{\as}$.%
\footnote{In the single-channel case ($f_q=0$)
  eq.~(\ref{eq:QplusGdot}) would imply that the sum rule violation is
  higher-twist only. In the matrix case, a higher twist component is expected
  on top of the perturbative component discussed here.}  This feature will be improved in
sec.~\ref{sec:knla} by modifying the
parameterization~(\ref{eq:K0param}) at NLO-NL$x$ level so as to
reduce the violation to $\order{\as^3}$.

\subsection{General form of the collinear $\boldsymbol\leftrightarrow$
  anti-collinear symmetry\label{ss:gfdrs}}

\begin{figure}[htbp]
  \centering
  \includegraphics[width=0.3\textwidth]{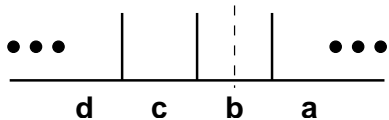}
  \caption{A sequence of splittings going towards smaller $x$ as $a\to
    b\to c \to d$, which may be collinear, anti-collinear or some
    mixture.}
  \label{fig:abcd}
\end{figure}

The role of the symmetry~(\ref{eq:symmetry}) in implementing RG properties
deserves a special discussion for the matrix kernel. In fact, its typical
effect is to produce two pole terms in $\ga$
and $1+\om-\ga$ (see eq.~(\ref{eq:chi_c})), which are supposed to
describe the correct product of $\Ga$-matrices for both direct and
reverse orderings. The situation is pretty clear for collinearly
ordered particles. In fact, referring to fig.~\ref{fig:abcd}, with $x$
decreasing from right to left we have a sequence of splitting
functions such as
\begin{equation}
  \label{eq:naive_collinear}
  \qquad\qquad\qquad\qquad
  \ldots \, \Ga_{dc} \,\Ga_{cb}\, \Ga_{ba} \, \ldots \qquad \qquad
  \begin{cases}
    x_d \,<\, x_c \,<\, x_b \,<\, x_a \\
    \kt_d \gg \kt_c \gg \kt_b \gg \kt_a
  \end{cases}
\end{equation}
describing $a$ going to $b$, $b$ to $c$ and so on, as predicted from
$\cK$ also.  On the other hand, in the anti-collinear limit the DGLAP
splitting functions need to account for the opposite splittings, $b$
to $a$, etc.
\begin{align}
  \label{eq:naive_anticollinear}
  \qquad\qquad\qquad\qquad
  &\; \ldots \, \Ga_{cd} \,\Ga_{bc}\, \Ga_{ab} \, \ldots \qquad \qquad
  \begin{cases}
    x_d \,>\, x_c \,>\, x_b \,>\, x_a \\
    \kt_d \ll \kt_c \ll \kt_b \ll \kt_a
  \end{cases} \\
  =&\; \ldots\, (\Ga^T)_{dc} \,(\Ga^T)_{cb}\, (\Ga^T)_{ba} \,\ldots \;.
  \nonumber
\end{align}

Then, one would naively expect that the anti-collinear pole in the
kernel be associated with $\Ga^T$. This seems to work fine as long as
we consider a complete chain of anti-collinear splittings.  Problems
arise however when trying to join collinear and anti-collinear chains.
Firstly there is an issue of colour factors: the anomalous dimension
$\Ga_{ij}$ implicitly includes a factor $N_i$ for the number of
varieties of parton $i$ that can be produced ($\NC^2-1$ if $i$ is a
gluon, $2 n_f\NC$ for an (anti)quark). For each exchanged particle in
fig.~\ref{fig:abcd} that factor should be included exactly once. In
the collinear limit it is included in the branching to the right of a
given exchange (e.g.\ for $b$ it is included in $\Ga_{ba}$), while in
the anti-collinear limit, as written in (\ref{eq:naive_anticollinear}),
it is included to the left (in $\Ga_{bc}$). If we are to consider a
single evolution from right to left containing both collinear and
anti-collinear splittings we should ensure that the $N_i$ factors are
consistently included to one side, for example in the branching to the
right of the exchange. One then needs to correct the splitting
function $(\Ga^T)_{ij}$ for an anti-collinear $i\to j$ splitting by a
factor $N_j/N_i$.


The second issue that arises relates to high-energy factorization. For
each exchanged gluon we have a factor $1/\om$. In a sequence of
collinear branchings that $1/\om$ factor is associated with the
splitting function to the right of the gluon exchange (e.g.\ if $b$ is
a gluon then it is included in $\Ga_{ba}$), while for anti-collinear
branchings it comes from the splitting function to the left (i.e.\
from $\Ga_{bc}$). This causes problems if we have an anti-collinear
splitting to the left of a gluon exchange and a collinear one to the
right, since \emph{both} will include a $1/\om$ factor for the
intermediate exchanged gluon. However it is necessary for the gluonic
part of our Green function to be consistent with high-energy
factorization, which systematically assigns an exchanged gluon's
$1/\om$ divergence to the larger-$x$ part of the diagram, i.e.\ to the
right of the exchanged gluon in the collinear limit. Therefore in the case of an
anti-collinear branching we should multiply $(\Ga^T)_{ij}$ by a factor
$f_{i}/f_{j}$ where $f_g = 1/\om$, so as to ensure that $j$ is never
associated with a $1/\om$ factor, while $i$ has it when $i$ is a
gluon.  Note that $f_q$ is arbitrary (other than that it should be a
non-zero constant for $\om \to 0$) since high-energy factorization is
not defined for quarks around $\om = 0$ --- we shall discuss its
choice below.

The outcome of this discussion is that colour factor and the high-energy
factorization corrections can be 
combined by introducing a similarity transformation matrix
\begin{equation}
  \label{eq:SimTrans}
  S = \left( 
    \begin{array}{cc}
      2\nf \NC f_q(\om)  & 0             \\ & \\
      0                  & (\NC^2-1) f_g(\om)
    \end{array}
  \right),
\end{equation}
and defining a `refactored' splitting function matrix
$\overline \Ga$ for anti-collinear splittings in an evolution that
will combine both collinear and anti-collinear splittings:
\begin{equation}
  \label{eq:UsingSimTrans}
  {\overline \Ga} = S \Ga^T S^{-1}
   = \left(
    \begin{array}{cc}
      \Ga_{qq}      &  \frac{n_f}{\CF}\frac{f_q(\om)}{f_g(\om)} \Ga_{gq} \\ & \\
       \frac{\CF}{n_f}\frac{f_g(\om)}{f_q(\om)} \Ga_{qg} & \Ga_{gg}
    \end{array}
   \right).
\end{equation}
A matrix kernel will therefore have collinear and anti-collinear
structure of the form
\begin{equation}
  \label{eq:basic_matrix_kernel}
  \cK \simeq \frac{\Ga}{\ga} + \frac{\overline \Ga}{1+\om-\ga} \; ,
\end{equation}
and will satisfy the collinear $\leftrightarrow$ anti-collinear symmetry in the general
form~(\ref{eq:genSymmetry}).  The fact that the diagonal entries (in
particular the $gg$ element) of $\Ga$ and ${\overline \Ga}$ are
identical is consistent with our expectation that the single-channel
($\nf=0$) limit should coincide with BFKL, which is symmetric in $\ga
\leftrightarrow 1+\om - \ga$.  The structure of colour factors and
$1/\om$ ensures that chains containing collinear and anti-collinear
splittings will have the expected sets of colour factors and overall
$1/\om$ factors.

Note finally that we can further specify $S$ so as to satisfy the
symmetry in the form~(\ref{eq:symmetry}). Since
\begin{equation}
  \label{eq:ga_qg__versus__ga_gq}
  \frac{\Ga_{qg,0}}{\Ga_{gq,0}} = \frac{2\nf \Tr}{\CF} \frac{\om}{\om+3}\,,
\end{equation}
we can simplify eq.~(\ref{eq:basic_matrix_kernel}) by exploiting the
arbitrariness of $f_q(\om)$ and setting it to
\begin{equation}
  \label{eq:effect_of_right_f_q}
  f_q(\om) = \frac{2\Tr}{\om+3} \quad \Longrightarrow \quad
  {\overline \Ga} = \Ga\;,
\end{equation}
thus providing, at leading level, a fully symmetric collinear structure
\begin{equation}
  \label{eq:basic_symmetric_matrix_kernel}
  \cK \simeq \Ga \left(  \frac1{\ga} + \frac1{1+\om-\ga} \right) \;,
\end{equation}
as assumed in eq.~(\ref{eq:symmetry}). 


\section{The kernel at NLO-NL$\boldsymbol x$ accuracy\label{sec:knla}}

\subsection{General structure of $\boldsymbol{\cK_0}$\label{ss:gsK0}}

Let us recall that, while the LO anomalous dimension matrix and the
LL$x$ expression of $\Ga_{gg}$ are factorization-scheme independent,
the NLO, NL$x$ expressions do depend on the scheme (except possibly
for the eigenvalue $\ga_+$ in the frozen $\as$ limit). This opens up
the possibility of constructing the kernel so as to reproduce the NLO,
NL$x$ anomalous dimensions in a given scheme, say $\MSbar$ scheme.
However, we have to comply with the
restrictions~(\ref{eq:gammaStruct}, \ref{eq:omegaStruct}, \ref{eq:symmetry}),
in particular the requirements of a simple $\ga$-pole structure at
fixed $\om$, a simple $\om$-pole at fixed $\ga$, and absence of
$1/\om$ singularity in $\cK_{qq}$ and $\cK_{qg}$. This means, at NLO,
that the $\as^2/\om$ terms of $\Ga_{qq}$ and $\Ga_{qg}$ cannot be
reproduced by an explicit $1/\om$ term in $\cK_1$, but should result
from the $\ga$-dependence of $\cK_{qq,0}$ and $\cK_{qg,0}$, where
$\cK_0^{(1)}, \; \cK_0^{(2)},\cdots$ are free scheme choice
parameters. In other words, we have to adjust the subleading
$\ga$-dependence of $\cK_0$ so as to reproduce the known $\MSbar$
anomalous dimensions at NL$x$ level in the form
\begin{equation}
  \label{eq:gamma1qg}
  \om \, \Ga_{qg,1} \sim (\cK_0^{(1)})_{qg}
  \quad \text{for} \quad \om \to 0\;.
\end{equation}

The above discussion shows that we have to change the
parameterization~(\ref{eq:K0param}) at next-to-leading level so as to
allow a more general subleading $\ga$-dependence. We choose the
following one
\begin{equation}
  \label{eq:chi_mat_0_all_prerequisites}
  \cK_0(\ga,\om) = 
    \begin{pmatrix}
      \Ga_{qq,0}(\om) \chi_c^\om(\ga) &
      \Ga_{qg,0}(\om) \chi_c^\om(\ga) + \Dqg(\om)\chiht^\om(\ga) \\ & \\
      \Ga_{gq,0}(\om) \chi_c^\om(\ga) & 
      \Ga_{gg,0}(\om) \chi_c^\om(\ga) + \displaystyle{\frac{2\CA}{\om}}
      \big[\chi_0^\om(\ga) - \chi_c^\om(\ga) \big]
    \end{pmatrix} \;,
\end{equation}
where $\chiht^\om(\ga)$ is a higher-twist kernel possessing the
symmetry~(\ref{eq:symmetry}), e.g.
\begin{equation}
  \label{eq:def_chi_ht}
  \chiht^\om(\ga) = \frac{2}{3}\left(
    \frac{1}{1+\ga} + \frac{1}{2+\om-\ga} \right) \;, \quad
  \chiht^0(0) = 1 \;,
\end{equation}
and $\Dqg$ is an $\om$-dependent coefficient which we require to be regular for
$\Re(\om) > -1$ and vanishingly small as $\om\to\infty$, e.g.%
\footnote{This particular choice is motivated by the fact that the $z$-space
  function $\Dqg(z) \sim (1-z)^2$ rapidly vanishes for $z \to 1$
  (cf.~eq.~(\ref{Dqgz})), thus not disturbing the large-$x$ behaviour of the
  model.}
\begin{equation}
 \label{Delta_qg}
 \Dqg(\om) \equiv \dqg \, \Delta(\om) \equiv \dqg \cdot 3
 \left( \frac{1}{1+\om} - \frac{2}{2+\om} + \frac{1}{3+\om} \right)
 \;, \quad \Dqg(0) = \dqg \;.
\end{equation}

The form~(\ref{eq:chi_mat_0_all_prerequisites}) allows one to choose
$(\cK_0^{(1)})_{qg}$ so as to reproduce the known NL$x$
expressions~(\ref{eq:Gamma0smallx}) and (\ref{eq:Gamma1smallx}) of
$\Ga_{qq}$ and $\Ga_{qg}$ in the $\MSbar$ scheme, up to order
$\as^2/\om$. Note that the logic is here reversed with respect to the
DIS scheme, in which $(\cK_0)_{qg}$ is directly calculated by
$\kt$-factorization (see~\cite{CaHa94}) and NL$x$ resummation formulae
for $\Ga_{qq}$ and $\Ga_{qg}$ are derived. By the $\ga \to 0$
behaviour
\begin{equation}
  \label{eq:gammaToZeroBehaviour}
  \cK_0 \to \frac1{\ga} \cK_0^{(0)} + \cK_0^{(1)} \; ,
\end{equation}
we derive, from eq.~(\ref{eq:Gamma_from_cK}) for $\Ga_1$, that
$\dqg^{\MSbar} = 8 \Tf / 9$.

In more detail, the NL$x$ coefficient $\zero\Ga_{qg,1}\equiv
\lim_{\om\to0}\,\om\,\Ga_{qg,1}$, according to eq.~(\ref{Gamma_1}) is given by
\begin{equation}
  \label{eq:gamma_qg1NLx}
  \zero\Ga_{qg,1}
  =\zero{\!\left[ \cK_1^{(0)} + \cK_0^{(1)} \cK_0^{(0)} \right]}_{qg}
  =\left[ \zero{\cK_1^{(0)}} + \zero{\cK_0^{(1)}} \; \one{\cK_0^{(0)}}
    + \one{\cK_0^{(1)}} \; \zero{\cK_0^{(0)}} \right]_{qg} \;.
\end{equation}
By expanding the matrix products in terms of their matrix elements and by taking
into account the conditions~(\ref{eq:quarkRowCond}), only the last term in the
r.h.s.\ of eq.~(\ref{eq:gamma_qg1NLx}) does not vanish, yielding
\begin{equation}
  \label{eq:gamma_qg1NLx_bis}
  \zero\Ga_{qg,1}
  = (\one{\cK_0^{(1)}})_{qg} \; (\zero{\cK_0^{(0)}})_{gg}
  = \left[ \Ga_{qg,0}(0) c_c(0) + \dqg \chiht^0(0) \right] 2\CA
  = 2\CA \left[ \frac43 \Tf + \dqg \right] \;,
\end{equation}
where we used the explicit expressions~(\ref{eq:def_chi_ht}),
(\ref{eq:constCoef}) and (\ref{c_c}).  The $\MSbar$-scheme value
$\zero\Ga_{qg,1} = 40 \Tf \CA / 9$ of eq.~(\ref{eq:Gamma1smallx}) is then
recovered provided
\begin{equation}
  \label{eq:Delta_qg_MSbar}
  \dqg^{\MSbar} = \frac{8 \Tf}{9} \;.
\end{equation}
Note that the corresponding expression for $\zero\Ga_{qq,1}$ involves
$(\zero{\cK_0^{(0)}})_{gq}=\frac{\CF}{\CA}(\zero{\cK_0^{(0)}})_{gg}$, thus
respecting the colour charge relation
$\Ga_{qq,n}^{\mathrm{NL}x}=\frac{\CF}{\CA}\Ga_{qg,n}^{\mathrm{NL}x}\;(n\geq1)$
which is apparent in eq.~(\ref{eq:Gamma1smallx}).

One could repeat the above procedure in other factorisation schemes as well. In
the $\DIS$ scheme, e.g., the value $\zero\Ga_{qg,1} = 52\Tf\CA / 9$ is recovered
by setting
\begin{equation}
  \label{eq:Delta_qg_DIS}
  \dqg^{\DIS} = \frac{14 \Tf}{9} \;.
\end{equation}
\subsection{BFKL limit and general structure of $\cK_1$\label{ss:gsK1}}

Having fixed $\Ga_{qq}$ and $\Ga_{qg}$ at NLO-NL$x$ level, the
remaining constraints (exact NLO anomalous dimension matrix and exact
BFKL kernel at NL$x$ level) are fixed by a proper choice of $\cK_1$.
Because of the expansion in eq.~(\ref{eq:omegaStruct})
\begin{equation}
  \label{eq:omegaExpan}
  \cK = \frac1{\om} \; \zero\cK(\as,\ga) + \one\cK(\as,\ga) + 
  \om \; \two{\cK}(\as,\ga) + \order{\om^2} \;,
\end{equation}
where $\zero\cK$ has only $gg$ and $gq$ entries,
the resolvent
\begin{equation}
  \label{eq:resolven}
  (1-\cK)^{-1} = (1-\one\cK)^{-1} \left[ 1 - \frac1{\om}\; \zero\cK
  (1-\one\cK)^{-1} + \order{\om} \right]^{-1}
\end{equation}
has the $gg$ matrix element proportional to a pure $\log 1/x$ evolution
form, with the kernel
\begin{align}
  K^{\mathrm{BFKL}} &= \asb K_0^\BFKL + \asb^2 K_1^\BFKL
  = \left[\zero\cK (1-\one\cK)^{-1} \right]_{gg}
  \nonumber \\
  &= \left[ \zero\cK + \zero\cK\;\one\cK \right]_{gg} + \order{\asb^3}\;,
  \label{eq:equivNLkern}
\end{align}
where $\asb \equiv\frac{\alpha_s N_c}{\pi}$.
We thus arrive at the identification
\begin{equation}
  \label{eq:NLx_ident}
  (2\CA) K_0^{\mathrm{BFKL}} = (\zero\cK_0)_{gg} \;, \quad
  (2\CA)^2 K_1^{\mathrm{BFKL}} = (\zero\cK_1
  + \zero\cK_0 \; \one\cK_0 )_{gg} \;,
\end{equation}
which parallels eq.~(\ref{eq:Gamma_from_cK}) for the perturbative
expansion, with the difference that it concerns the $gg$ entry only,
as is appropriate to the $\kt$-factorization of gluon exchange.
Eq.~(\ref{eq:NLx_ident}) is used --- as in the single-channel
case~\cite{rgiggf} --- to derive $(\zero\cK_1)_{gg}$ from the known
expression of the NL$x$ BFKL kernel
\footnote{Of course, running coupling contributions to $K_1^\BFKL$ ---
  explicitly considered in sec.~\ref{ss:runningc} --- are to be subtracted out.}
$K_1^\BFKL$ and of the $\cK_0$ kernel~(\ref{eq:chi_mat_0_all_prerequisites}).
Explicitly
\begin{align}
  \label{eq:NLsubtr}
  ( \zero\cK_0 \; \one\cK_0 )_{gg}
  &= 2\CA \chi_0 ( 2\CA \dot{\chi}_0 + A_{gg} \chi_c )
    + 2\CF \chi_c ( \Ga_{qg,0} \chi_c + \dqg \chiht )\Big|_{\om=0}
 \\ \label{chi0dot}
 \dot{\chi}_0(\ga) &\equiv \partial_\om \chi_0^\om(\ga)\big|_{\om=0} = -\psi'(1-\ga) \;,
\end{align}
where $A_{gg}(\om) \equiv \Ga_{gg,0}(\om) - 2\CA/\om$ is the regular
part of the $gg$ anomalous dimension.  This result reduces to the
corresponding one-channel~\cite{rgiggf} $\nf=0$ subtraction
$2\CA\chi_0(2\CA\dot\chi_0+A_{gg}^{[\nf=0]}\chi_c)$ after setting
$\Ga_{qg} = \Dqg = 0$.

We encounter at this point a consistency relation on the factorization scheme
for $\Ga_1$, due to the fact that we want to incorporate both $\Ga_1$
and $K_1^\BFKL$ in a kernel satisfying the simple-pole assumptions of
sec.~\ref{ss:bckc}, as better discussed in sec.~\ref{ss:cr}. Note in fact that, given
$\cK_0$, eq~(\ref{eq:Gamma_from_cK}) determines  $\cK_1^{(0)}$ in terms of
$\Ga_1$  and eq.
(\ref{eq:NLx_ident}) determines $(\zero\cK_1)_{gg}$ in terms of $K_1^\BFKL$. Therefore,
$\big[\zero\cK_1^{(0)}\big]_{gg}$ --- the $\ga$-pole part with the $1/\om$
singularity --- is determined in two independent ways, which should provide
the {\em same} result, in the form
\begin{equation}
  \zero{\!\big[\Ga_{gg,1} - (\cK_0^{(1)}\cK_0^{(0)})_{gg} \big]}
  = \zero{\!\left( \ga_{+,1}\right)} - (\zero\cK_0\;\one\cK_0)_{gg}^{(0)} \;,
  \label{eq:ggConsCons}
\end{equation}
where we have used the fact that the simple-pole part of $K_1^\BFKL$ predicts
the NL$x$ part of $\ga_{+,1}$, that is
\begin{equation}
  \label{eq:zeroGammaPlus}
  \zero{\!\left(\ga_{+,1}\right)} = (2\CA)^2 \left( K_1^\BFKL \right)^{(0)} \;.
\end{equation}
We show in sec.~(\ref{ss:cr}) that the consistency
equation~(\ref{eq:ggConsCons}) is identically satisfied, provided
$\zero\Ga_{qg,1} \simeq \big[\cK_0^{(1)}\big]_{qg}$, as assumed in
eq.~(\ref{eq:gamma1qg}). Therefore, both DIS and $\MSbar$ schemes satisfy
eq.~(\ref{eq:ggConsCons}) and can be accommodated by a proper matrix kernel at
NL level.

More precisely, we start from the ``Ansatz''
\begin{equation}
  \label{eq:our_K1_struct}
  \cK_1(\ga,\om) = \widetilde{\Ga}_1(\om) \,\chi_c^\om(\ga)
   + (2\CA)^2\left(\frac1{\om}-\frac2{1+\om}\right)
    \begin{pmatrix}
      0 &  0\\
      0 &  \tilde{\chi}_1^\om(\ga)
    \end{pmatrix} \;.
\end{equation}
where\\
{\it (i)} the function $\tilde{\chi}_1^\om(\ga)$ at $\om = 0$ is equal to
\begin{equation}
 \label{chiTilde}
 \tilde{\chi}_1^{\om=0} \equiv \tilde{\chi}_1 = \frac{\zero\cK_{gg,1}}{(2\CA)^2}
 = K_1^\BFKL - \frac{\big[\zero\cK_0 \; \one\cK_0\big]_{gg}}{(2\CA)^2}
\end{equation}
[see eq.~(\ref{chi1tilde}) for an explicit expression] and is extrapolated to
generic $\om$ values by the $\om$-shift~\cite{rgiggf} procedure of left and right
projections as follows
\begin{equation}
  \label{eq:chi1tildeom}
  \tilde{\chi}_1^\om(\ga) = [\tilde{\chi}_1]_L(\ga) +
  [\tilde{\chi}_1]_L(1+\om-\ga) \;,
\end{equation}
so as to satisfy the symmetry~(\ref{eq:symmetry});\\
{\it (ii)} in order to minimize momentum sum rule violations, we have added to
the high-energy pole $1/\om$ in front of
$\tilde{\chi}_1^\om$ a low-energy term $-2/(1+\om)$: their sum vanishes at $\om=1$;\\
{\it (iii)} the $\widetilde{\Ga}_1(\om)$ matrix is fixed by matching
$\cK_1^{(0)} + \cK_0^{(1)}\cK_0^{(0)}$ to the known NLO $\MSbar$ splitting
functions (cf.~eq.~(\ref{Gamma_1})):
\begin{equation}
  \widetilde{\Ga}_1(\om)  = {\Ga}_1^{(\MSbar)}(\om) 
  - \cK_0^{(1)}\cK_0^{(0)} - (2\CA)^2\left(\frac1{\om}-\frac2{1+\om}\right)
  \begin{pmatrix}
    0 &  0\\
    0 & \tilde{\chi}_1^{(0)}
  \end{pmatrix} \;,
\end{equation}
$\tilde{\chi}_1^{(0)}$ being the coefficient of the simple pole at $\ga=0$ of
$\tilde{\chi}_1^\om(\ga)$. We note that $\chi_c^\om$ in
eq.~(\ref{eq:our_K1_struct}) is again chosen for simplicity reasons according to
the symmetry~(\ref{eq:symmetry}).

The final form of the next-to-leading matrix kernel is then
\begin{equation}
  \label{eq:K1ansatz}
  \cK_1 = \left(\Ga_1 - \cK_0^{(1)} \cK_0^{(0)} \right) \chi_c^{\om}
  + (2\CA)^2\left(\frac1{\om}-\frac2{1+\om}\right)
  \begin{pmatrix}
    0 & 0 \\ 0 & \tilde{\chi}_1^\om - \tilde{\chi}_1^{(0)} \chi_c^\om
  \end{pmatrix} \;.
\end{equation}

\subsection{Running coupling features\label{ss:runningc}}

We shall choose the running coupling scales of our kernel as in~\cite{rgiggf}.
We thus associate $\as(q^2)$ ($\qt \equiv \kt-\kt'$) to the LL$x$ BFKL kernel
$\chi_0^{\om}(\ga)$, and $\as(k_>^2)$ ($k_> \equiv \max\{k,k'\}$) to all other
ones. The choice of the intermediate gluon momentum transfer is suggested by the
NL$x$ BFKL kernel itself, which contains the beta-function dependent
term 
\begin{equation}\label{chirun}
  \chi^\run(\ga) = -\frac{b}{2} \left( \chi_0'+\chi_0^2 \right) \;,
  \qquad b = \frac{11}{12} - \frac{\Tf}{3\CA}\,,
\end{equation}
(quoted for a renormalisation scale choice $\mu^2 = k^2$), corresponding to the $\kt$-space kernel
\begin{equation}\label{Krun}
  K^\run(\kt,\kt') = -b \left[ \log\frac{q^2}{k^2} K_0^\BFKL(\kt,\kt')\right]_{\mathrm{reg}}
\end{equation}
where the regularization procedure is explained in~\cite{rgiggf}. Since the
term in eq.~(\ref{Krun}) is accounted for by expanding $\as(q^2)$ up to NL order,
the expression in eq.~(\ref{chirun}) should be subtracted out from the NL kernel
considered before. More precisely, the kernel $K_1^\BFKL$ in eq.~(\ref{eq:NLx_ident}) is
meant to have the eigenvalue
\begin{equation}\label{chi1BFKL}
 \chi_1^\BFKL = \left[\chi_1 -\frac12 \chi_0 \chi_0' \right]
 + \frac{b}{2} \left( \chi_0' + \chi_0^2 \right) \;,
\end{equation}
where the expression in square brackets is the eigenvalue function at energy-scale $k^2$,
obtained by the $\om$-expansion of eq.~(\ref{eigenvChange}), and $\chi_1$ is the
customary NL eigenvalue~\cite{FaLi98,CaCi98} at energy-scale $k k_0$, given by the expression
\begin{align}\nonumber
 \chi_1(\ga) &= -\frac{b}{2} [\chi^2_0(\ga) + \chi'_0(\ga)]
  -\frac{1}{4} \chi_0''(\ga)
  -\frac{1}{4} \left(\frac{\pi}{\sin \pi \ga} \right)^2
  \frac{\cos \pi \ga}{1-2\ga}
  \left[3+\left(1+\frac{2\Tf}{\CA^3}\right)\frac{2+3\ga(1-\ga )}{(1+2\ga)(3-2\ga)}\right] \\
 &\quad +\left(\frac{67}{36}-\frac{\pi^2}{12} -\frac{5\Tf}{9\CA} \right) \chi_0(\ga)
  +\frac{3}{2} \zeta(3) + \frac{\pi^3}{4\sin \pi\ga}  \nonumber\\
 &\quad - \sum_{n=0}^{\infty} (-1)^n
 \left[ \frac{\psi(n+1+\ga)-\psi(1)}{(n+\ga)^2}
 +\frac{\psi(n+2-\ga)-\psi(1)}{(n+1-\ga)^2} \right] \; .
\label{eq:nllorg}
\end{align}

It follows that the overall kernel has the structure
\footnote{Its
  generalisation to include variable renormalisation scale is
  constructed as follows: single powers of $\as$ undergo the
  transformation $\alh(q^2) \to \alh(x_\mu^2 q^2) + \beta_0
  \alh^2(x_\mu^2 q^2) \ln x_{\mu}^2$ and $\alh(k_>^2) \to \alh(x_\mu^2
  k_>^2) + \beta_0 \alh^2(x_\mu^2 k_>^2) \ln x_{\mu}^2$, with $\beta_0
  = 2 \CA b$, while
  quadratic powers of $\as$ are modified as $\alh^2(k_>^2) \to
  \alh^2(x_\mu^2 k_>^2)$.}
\begin{align} \label{runningKernel}
 \cK(\kt,\kt';\om) &= \alh(q^2) \frac{2\CA}{\om}
 \begin{pmatrix}
   0 & 0 \\[1ex] 0 & K_0^\om
 \end{pmatrix}
 + \alh(k_>^2)
 \begin{pmatrix}
  \Ga_{qq,0} K_c^\om & \Ga_{qg,0} K_c^\om + \Dqg \Kht^\om \\[1ex]
  \Ga_{gq,0} K_c^\om & \left(\Ga_{gg,0}-\frac{2\CA}{\om}\right) K_c^\om
 \end{pmatrix}
  \\[1ex] \nonumber
 &\quad + \alh^2(k_>^2) \left[ \left(\Ga_1-\cK_0^{(1)}\cK_0^{(0)}\right) K_c^\om
 + (2\CA)^2\left(\frac1{\om}-\frac2{1+\om}\right)
 \begin{pmatrix}
   0 & 0 \\[1ex] 0 & \tilde{K}_1^\om - \tilde{\chi}_1^{(0)} K_c^\om
 \end{pmatrix} \right] \;.
\end{align}
$K_0^\om$, $K_c^\om$ and $\Kht^\om$ are the $\kt$-dependent kernels corresponding
to the characteristic functions $\chi_0^\om$, $\chi_c^\om$ and $\chiht^\om$
respectively, and their explicit expressions in $(x,k)$-space are provided in
app.~\ref{a:kcf} (cf.~eqs.~(\ref{K0action}), (\ref{K_c_bis}) and (\ref{K_ht})).
The eigenvalue of the $gg$ entry of $\cK_1$ at $\om = 0$, thanks to
eq.~(\ref{chi1BFKL}) and to the subtraction procedure in eq.~(\ref{eq:NLsubtr}),
is provided by the expression
\begin{align}
 \tilde{\chi}_1^{\om=0}(\ga) &= \chi_1(\ga)
 + \frac{b}{2} \left[ \chi_0'(\ga) + \chi_0^2(\ga) \right]
 + \frac12 \chi_0(\ga) \frac{\pi^2}{\sin^2(\pi\ga)} \nonumber \\
 &- \chi_0(\ga) \frac{A_{gg}(0)/2\CA}{\ga(1-\ga)}
 -\frac{\CF/\CA}{\ga(1-\ga)} \left[ \frac{\Ga_{qg,0}(0)/2\CA}{\ga(1-\ga)}
 + \frac{\dqg/\CA}{(1+\ga)(2-\ga)} \right] \;,
 \label{chi1tilde}
\end{align}
which then acquires the $\om$-dependent shift, as explained previously.  Note
that $\tilde{\chi}_1$ is the same as that in~\cite{rgiggf} in the $n_f=0$ limit,
and differs from it by the $n_f$-dependent terms in $A_{gg}$, $\Ga_{qg,0}$ and
$\Delta_{qg}$. Note also that cubic and quadratic poles cancel out in
$\tilde{\chi}_1$, because of the $\om$-shift of the collinear poles and of their
factorization, which are embodied in our formalism (cf.~sec~\ref{ss:cr}).

The remaining single poles of $\cK_{gg,1}$ are provided by $\tilde{\chi}_1^{(0)}$, which
is obtained as follows. Let us introduce the constant (in $\ga$) coefficients
$c_{\chi}$ of the characteristic functions:
\begin{equation}
  \label{eq:constCoef}
  \chi_0(\ga,\om) \equiv \frac1{\ga} + c_0(\om) + \order{\ga} \;, \quad
  \chi_c(\ga,\om) \equiv \frac1{\ga} + c_c(\om) + \order{\ga} \;.
\end{equation}
Then, according to eqs.~(\ref{eq:chi_0},\ref{eq:chi_c}), we have
\begin{subequations}\label{constants}
  \begin{align}
    c_0(\om) &= \psi(1)-\psi(1+\om) \;, \quad c_0(0) = 0 \;, \quad c'_0(0) =
    -\psi'(1) = -\frac{\pi^2}{6}
    \label{c_0} \\
    c_c(\om) &= \frac1{1+\om} \;, \quad c_c(0) = 1
    \label{c_c}
  \end{align}
\end{subequations}
Note that $c_0(0)$ vanishes by virtue of the LL$x$ expansion $\chi_0(\ga) =
1/\ga + \order{\ga^2}$, in other words it is a scheme-independent
coefficient. On the other hand, all other quantities in eq.~(\ref{constants})
are scheme-dependent, i.e., they depend on the particular choice we adopted for
shifting the poles of $\chi_0$ and on the definition of $\chi_c$ as in
eqs.~(\ref{eq:chi_0}, \ref{eq:chi_0}).

By using the above notation we derive the expansions of the BFKL eigenvalue
function
\begin{equation}\label{chi1gaExpn}
  \chi_1^\BFKL(\ga) =
  \frac1{\ga^2}\left(\frac{A_{gg}}{2\CA}
   + \frac{\CF}{\CA}\frac{\Ga_{qg,0}}{2\CA}\right)
  -\frac1{\ga} \frac{(46\CA-52\CF)\Tf}{9(2\CA)^2}
  +\order{\ga^0} \;,
\end{equation}
the expansion of the subtraction term in eq.~(\ref{eq:NLsubtr})
\begin{equation}\label{subExpn}
  (\ref{eq:NLsubtr}) =
  \frac1{\ga^2}\left(\frac{A_{gg}}{2\CA} + \frac{\CF}{\CA}\frac{\Ga_{qg,0}}{2\CA}\right)
  + \frac1{\ga}\left[c'_0
    + c_c\left(\frac{A_{gg}}{2\CA} + \frac{\CF}{\CA}\frac{2\Ga_{qg,0}}{2\CA} \right)
    + \frac{\CF}{\CA}\frac{\dqg}{2\CA} \right] +\order{\ga^0}
  \end{equation}
and the expression of the pole term in $\cK_1$
\begin{equation}\label{chi1t0}
  \tilde{\chi}_1^{(0)} =
  -\frac{(46\CA-52\CF)\Tf}{9(2\CA)^2}
  + c_c\left( \frac{11}{12}
  + \frac{(4\CA-16\CF)\Tf}{3(2\CA)^2}
  \right) - \frac{\CF}{\CA}\frac{\dqg}{2\CA} \;,
\end{equation}
where all quantities in the three formulas above are evaluated at $\om = 0$.
Note again that $\tilde{\chi}_1^{(0)}$ satisfies the consistency relation in
eq.~(\ref{eq:ggConsCons}), proved in sec.~\ref{ss:cr}.

The remaining expressions used in the previous subsections are easily obtained
by the following detailed formulas
\begin{subequations}\label{K01K00ij}
  \begin{align}
    [\cK_0^{(1)} \cK_0^{(0)}]_{qq} &= c_c \left(\Ga_{qg} \Ga_{gq} + \Ga_{qq}^2
    \right) + \Dqg \Ga_{gq} \chiht && \sim \Ga_{qq,1} \sim
    \frac{\CF}{\CA}\Ga_{qg,1}
    \\
    [\cK_0^{(1)} \cK_0^{(0)}]_{qg} &= c_c \Ga_{qg}
    \left(\Ga_{qq}+\Ga_{gg}\right) + \Dqg \Ga_{gg} \chiht && \sim \Ga_{qg,1}
    \\
    [\cK_0^{(1)} \cK_0^{(0)}]_{gq} &= \Ga_{gq} \left[ c_0
      \, \frac{2\CA}{\om} + c_c \left( \Ga_{qq} + A_{gg} \right) \right]
    \\
    [\cK_0^{(1)} \cK_0^{(0)}]_{gg} &= c_0 \Ga_{gg} \, \frac{2\CA}{\om}
      + c_c \left[ A_{gg} \Ga_{gg} + \Ga_{gq} \Ga_{qg} \right]
   && \sim \Ga_{gg,1} - (2\CA)^2 \frac{\tilde{\chi}_1^{(0)}}{\om} \;,
  \end{align}
\end{subequations}
where $\chiht$ stands for $\chiht^\om(\ga=0)$ and $\sim$ means asymptotic in the
$\om \to 0$ limit. From the above formulas we can compute the high-energy limit
of $\cK_1$
\begin{equation}\label{K1heLim}
 \lim_{\om \to 0} \om\cK_1 \equiv \zero\cK_1 = (2\CA)^2
  \begin{pmatrix}
   0 & 0 \\ \kappa_{gq}\chi_c(\ga) & \tilde{\chi}_1(\ga)
  \end{pmatrix}
 \;, \qquad \kappa_{gq} \equiv \textstyle{ \frac{\CF}{\CA}\left(\frac14 -\frac{10\Tf}{9\CA} - c_0'
     - c_c\frac{A_{gg}}{2\CA}\right)} \;.
\end{equation}

The introduction of running coupling may change our expectations on how momentum
conservation is satisfied by our kernel. Since we incorporate NLO anomalous
dimensions, as given by eqs.~(\ref{eq:K1ansatz}), we expect violations at next
order, that is at relative order $\as^3$. This is also the order at which
running coupling effects start to matter in the actual derivation of anomalous
dimensions, being related to a commutator of two values of $\as$, evaluated at
different scales (cf.\ app.~\ref{a:read}). Therefore, no problems arise at NLO.
Since we do not explicitly consider incorporating NNLO results in this paper, we
do not make an effort to improve energy-momentum conservation at frozen $\as$.
We should add that the consistency relations on NL$x$ terms which restrict our
scheme start imposing a collinear scheme restriction at order $\as^3/\om^2$
which is derived in sec.~\ref{ss:rf} and is violated, even if marginally, in
the $\MSbar$-scheme.

\section{Frozen coupling anomalous dimensions\label{s:fcad}}

In this section we want to discuss a number of issues concerning the anomalous
dimension matrix in the case of frozen coupling, in which the whole matrix can
be analytically calculated in terms of the kernel matrix elements in
$(\ga,\om)$-space. This allows us to compute the two eigenvalues
$\ga=\ga_{\pm}(\as,\om)$ and their inverses, the effective eigenvalue
functions $\om=\chi_{\pm}(\as,\ga)$, as well as their eigenvectors. We
obtain in this way the hard Pomeron exponent $\oms(\as)$ and the resummation
formulae for the matrix elements of the anomalous dimension matrix. The
latter, at a given level of the LL$x$ hierarchy, must be consistent with the
exact low order anomalous dimensions we have used in constructing the kernel,
thus providing consistency relations for the collinear and $\kt$-factorization
schemes. Here we find, at NL$x$ level, that such relations are identically
satisfied by our construction at NLO, while they put a nontrivial constraint
on the $\as^3/\om^2$ term of $\Ga_{gq}$ at NNLO.

\subsection{Anomalous dimension matrix\label{ss:adm}}

If $\as$ is frozen, the matrix kernel $\cK(\kt,\kt';\om)$ is scale invarant
and its resolvent admits the $\ga$-representation
\begin{equation}\label{eq:resolvent}
  \G(\kt,\kt_0;\om) = \frac1{k^2}\int\frac{\dif\ga}{2\pi\ui}
  \left(\frac{k^2}{k_0^2}\right)^\ga \frac1{1-K(\as,\ga,\om)} \;,
\end{equation}
where $\cK(\as,\ga,\om)$ is the characteristic function matrix of
$\cK(\kt,\kt';\om)$.

We introduce the eigenvalues $\eta_\pm$ and eigenvectors $u_\pm$ of $\cK$ in the
usual way
\begin{equation}\label{eq:eigenv}
  \cK(\ga) u_\pm(\ga) = \eta_\pm(\ga) u_\pm(\ga) \;,
\end{equation}
where the $\as$- and $\om$-dependences of all the above quantities are understood.
One can then write the spectral decomposition
\begin{equation}\label{eq:specDec}
  \cK(\ga) = \eta_+(\ga) \Pi_+(\ga) + \eta_-(\ga) \Pi_-(\ga) \;,
\end{equation}
where $\Pi_\pm$ are the orthogonal projectors on the eigenspaces of $\cK$, and
are given by
\begin{equation}\label{eq:projectors}
  \Pi_+ = \frac{u_+ \otimes \bar{v}_+}{(\bar{v}_+ u_+)} \;, \quad
  \Pi_- = \frac{u_- \otimes \bar{v}_-}{(\bar{v}_- u_-)} \;,
\end{equation}
$\bar{v}_\pm(\ga)$ being the left-eigenvectors of $\cK$ satisfying
$(\bar{v}_+ u_-) = 0 = (\bar{v}_- u_+)$ when $u$'s and $\bar{v}$'s are evaluated
at the same value of $\ga$.

The behaviour in $t$-space of $\G$ is determined by the $\ga$-poles in
eq.~(\ref{eq:resolvent}). These poles are found at $\ga = \ga_\pm(\as,\om)$ such
that
\begin{equation}\label{eq:gammaPoles}
  \eta_+(\ga_+,\as,\om) = 1 \;, \quad \eta_-(\ga_-,\as,\om) = 1
\end{equation}
and are interpreted as anomalous dimension eigenvalues of $\G$.

By applying a driving term $f_0$ to the Green's function $\G$, at leading-twist
level --- i.e., taking into account only the two rightmost poles in the
half-plane $\Re(\ga) < 1/2$ obeying $\lim_{\as\to0}\ga_\pm \to 0$ --- one
obtains the vector of (integrated) quark and gluon densities
\begin{equation}\label{eq:Gf0}
  \column{f_q}{f_g}  \equiv f 
= \sum_{l\in\{+,-\}} \frac1{-\ga_l \,\eta_l'(\ga_l)}
 \left(\frac{k^2}{k_0^2}\right)^{\ga_l}
 \Pi_l(\ga_l) f_0(\ga_l)
\end{equation}
We want to show that $f$ satisfies the DGLAP-type evolution equation
\begin{equation}\label{eq:dglap-type}
  \frac{\dif f}{\dif \log k^2} = \Ga f \;,
\end{equation}
in terms of a well-defined resummed anomalous dimension matrix $\Ga$. In
fact, by inserting the expression~(\ref{eq:Gf0}) into both sides of
eq.~(\ref{eq:dglap-type}), the equality is satisfied provided
\begin{equation}\label{eq:GammaCondition}
  [\Ga-\ga_l\mathbb{I}] \Pi_l(\ga_l) f_0(\ga_l) = 0 \;, \qquad(l = +,-) \;.
\end{equation}
It might seem that $\Ga$ is dependent on the initial condition $f_0$. This
is not the case, because whatever the choice of $f_0$, the projector
$\Pi_l(\ga_l)$ projects $f_0$ into a vector proportional to
$u_l(\ga_l)$, and eq.~(\ref{eq:GammaCondition}) reduces to
\begin{equation}\label{eq:GammaCondition2}
  \Ga u_l(\ga_l) = \ga_l u_l(\ga_l) \;, \qquad(l = +,-) \;,
\end{equation}
i.e., $\Ga$ is the (unique) matrix whose eigenvectors are
$\{u_+(\ga_+),u_-(\ga_-)\}$ relative to the eigenvalues $\{\ga_+,\ga_-\}$.
\begin{equation}\label{eq:Gamma}
 \Ga = \ga_+ \frac{u_+(\ga_+) \otimes \bar{v}_+(\ga_-)}
 {\bar{v}_+(\ga_-) \cdot u_+(\ga_+)}
 + \ga_- \frac{u_-(\ga_-) \otimes \bar{v}_-(\ga_+)}
  {\bar{v}_-(\ga_+) \cdot u_-(\ga_-)} \;.
\end{equation}
In more detail, the eigenvalues $\ga_\pm(\as,\om)$ are provided by
\begin{equation}\label{eq:det1mK}
 \det[1-\cK(\ga_\pm,\as,\om)] = 0
\end{equation}
and the eigenvectors are
\begin{align}
 u_+(\ga_+) &= \column{\rho}{1} \;, &
 \rho &\equiv \frac{\cK_{qg}(\ga_+)}{1-\cK_{qq}(\ga_+)}
 = \frac{1-\cK_{gg}(\ga_+)}{\cK_{gq}(\ga_+)}
 \label{uplus} \\
 u_-(\ga_-) &= \column{1}{-r} \;, &
 r &\equiv  \frac{\cK_{gq}(\ga_-)}{\cK_{gg}(\ga_-)-1}
 = \frac{\cK_{qq}(\ga_-)-1}{\cK_{qg}(\ga_-)}
 \label{uminus} \\
 \bar{v}_-(\ga_+) &= \row{1}{-\rho} \;, &&
 \label{vplus} \\
 \bar{v}_+(\ga_-) &= \row{r}{1} \;. && 
 \label{vminus} 
\end{align}
Therefore, the {\em full} expression of the anomalous dimension matrix is
\begin{align}
 \Ga &= \frac{\ga_+}{1+r\rho} \column{\rho}{1}\otimes\row{r}{1}
 + \frac{\ga_-}{1+r\rho} \column{1}{-r}\otimes\row{1}{-\rho}
 \label{GammaSpectral} \\ \nonumber
 &= \frac1{1+r\rho}
 \begin{pmatrix}
   r\rho\ga_+ + \ga_- & (\ga_+ - \ga_-) \rho \\[1ex]
   (\ga_+ - \ga_-) r  & \ga_+ + r\rho\ga_-
 \end{pmatrix} \;.
\end{align}
We obtain the relationships
\begin{align}
 \ga_+ &= \Ga_{gg} + r \Ga_{qg} \;,
 \label{gammaPlus} \\
 \ga_- &= \Ga_{qq} - r \Ga_{qg} \;,
 \label{gammaMinus} \\
 \Ga_{gq} &= r (\Ga_{gg} - \ga_-) \;.
 \label{gammaGiQu}
\end{align}

\subsection{Resummation formulae\label{ss:rf}}

All the above formulas are exact in the frozen coupling case, and do not depend
on our particular assumptions on $\cK$. Now, by taking into account the
structure of $\cK$ described in the previous sections, we compute the anomalous
dimension matrix elements at NL$x$ level. 

To this purpose, we note that the eigenvalues of $\cK$ are defined by
\begin{equation}
  \eta^2 - \eta \tr \cK + \det \cK = 0
\end{equation}
and that both $\tr\cK$ and $\det\cK$ are of order $1/\om$, so that up to NL$x$
level we have
\begin{align}
 \eta_+ &\simeq \tr \cK - \frac{\det \cK}{\tr \cK} \simeq \frac{\alh}{\om}
  \left(\zero\cK_{gg,0} + \alh \zero\cK_{gg,1} \right) + \alh \left(
  \one\cK_{gg,0} + \frac{\zero\cK_{gq,0} \; \one\cK_{qg,0}}{\zero\cK_{gg,0}}
  \right) \;, \\
 \eta_- &\simeq \frac{\det \cK}{\tr \cK} \simeq \alh \left( \one\cK_{qq,0}
  - \frac{\zero\cK_{gq,0} \; \one\cK_{qg,0}}{\zero\cK_{gg,0}} \right) \;.
\end{align}
We note that the equation $\eta_+=1$ reduces to the usual BFKL determination of
$\ga_+$ because $\zero\cK_{gg,0}$ can be replaced by
$\om/\alh$ in the NL$x$ term. Furthermore, the equation
$\eta_-=1$ is dominated by its $\ga$-pole part, yielding
\begin{equation}
 \eta_- \simeq
 \frac{\alh}{\ga} \left( \one\Ga_{qq,0} - \frac{\CF}{\CA} \, \one\Ga_{qg,0} \right)
 = -\frac{\CF}{\CA} \frac{\Ga_{qg,0}(\om=0)}{\ga} = 1 \;.
\end{equation}
This provides the lowest order determination of
\begin{equation}\label{eq:gamma-NL}
 \ga_- = \frac{\as}{2\pi} \ga_{-,0} + \text{NNL}x
  = - \frac{\CF}{\CA} \Ga_{qg,0}(0) + \text{NNL}x \;,
\end{equation}
so that, up to NL$x$ level only the one-loop term of $\Ga_{qg}$ contributes
and no small-$x$ enhancements are present.

The coefficient $r$ is now calculable from eq.~(\ref{uminus}) and, up to NL$x$
level, we obtain
\begin{equation}\label{eq:r}
 r = \frac{\CF}{\CA}\left[ 1 + \frac{\as}{2\pi} r_1 + \om \tilde{r}_1
 + \cdots\right] \;, \quad
 r_1 = c_c \ga_{-,0} + \frac{\cK_{gq,1}^{(0)}}{2\CF}
 - \frac{\cK_{gg,1}^{(0)}}{2\CA} \;, \quad
 \tilde{r}_1 = \frac{A_{gq}}{2\CF} - \frac{A_{gg}+\frac{\CF}{\CA} \Ga_{qg,0}}{2\CA} \;.
\end{equation}
Note that, since eq.(\ref{uminus}) is
evaluated at $\ga_-$, $r$ does not contain $1/\om$ enhancements and is
generally calculable  from fixed order perturbation theory. The coefficient $\rho$, on the
other hand, is calculated at $\ga_+$ (cf.~eq.~(\ref{uplus})), and so contains
resummation of NL$x$ terms to all orders in $\as$.

Given that $\ga_-$ and $\rho$ are NL$x$ quantities, it follows that also
$\Ga_{qq}$ and $\Ga_{qg}$ are NL$x$. From the previous equations we obtain the
resummation formulae
\begin{align}
 \Ga_{gq}^{\LL} &= \frac{\CF}{\CA} \Ga_{gg}^{\LL}
  = \frac{\CF}{\CA} \ga_+^{\LL}
 \label{eq:gq_gg_LL_rel} \\
 \Ga_{qg}^{\NL} &= \ga_+^{\LL} \rho = \ga_+^{\LL} \cK_{qg}(\ga_+)
 \label{gaqgNL} \\
 \Ga_{qq}^{\NL} &= \frac{\CF}{\CA}
 \left( \Ga_{qg}^{\NL} - \Ga_{qg,0}(0) \right) \;,
 \label{gaqqNL}
\end{align}
which are well known~\cite{CaHa94}. In addition, the matrix kernel predicts
\begin{align}
 \Ga_{gq}^{\NL} &= \frac{\CF}{\CA} \left[ \ga_+^{\NL} - \Ga_{qq}^{\NL}
 + \frac{\as}{2\pi} r_1 \ga_+^{\LL} + \om \tilde{r}_1 \ga_+^{\LL} \right]
 \label{gagqNL} \\ \nonumber
 &= \frac{\CF}{\CA} \left[ \Ga_{gg}^{\NL} + \frac{\as}{2\pi}\frac{\CF}{\CA}
 \Ga_{qg,0}(0) + \frac{\as}{2\pi} \frac{\asb}{\om} r_1 + \asb \tilde{r}_1 \right]
 + \order{\as^4} .
\end{align}
Note that NL$x$ running coupling contributions are shown in Appendix A to
start at order $\as^4$. The above resummation
formula for $\Ga_{gq}$
is easily checked to be identically valid at $\order{\as}$ and
$\order{\as^2}$. At $\order{\as^3}$ it yields the relation
\begin{equation}\label{eq:NNLOconstraint}
 \Ga_{gq,2}^{\NL} = \frac{\CF}{\CA} \Ga_{gg,2}^{\NL} \;,
\end{equation}
which characterises the class of schemes described by our matrix formulation,
and appears to be not satisfied in the $\MSbar$ scheme~\cite{NNLO}
\footnote{From eq.~(4.29) of ref.~\cite{NNLO}, by taking into account the
  difference between our and their normalization
  $\Ga_{ab,2}^{\NL} = -E^{ab}_1 / 8\om^2$, it turns out that, in the
  $\MSbar$-scheme,
  $\Ga_{gq,2}^{\NL} = \frac{\CF}{\CA} \left[ \Ga_{gg,2}^{\NL}
  -\frac{\nf}{3\om^2} \right]$.
}%
, even though the violation, of relative order $n_f/N_c^2$, is numerically less
than 0.5\%  for $\nf \leq 6$. Strictly speaking, this implies that the $\MSbar$
scheme at NNLO cannot be incorporated in the present matrix approach. However,
one could think of adding the small violation just mentioned by a matching
procedure.  

\subsection{Consistency relations\label{ss:cr}}

They arise in general because of the joined requirements of simple $\om$-poles
and $\ga$-poles inposed on our kernel. For instance, by the $\ga$-pole
hypothesis we determine the $\ga$-pole parts of $\cK_1$ and $\cK_2$ by the
equations
\begin{equation}
  \label{gammaPolePart}
  \cK_1^{(0)} = \Ga_1 - \cK_0^{(1)} \Ga_0 \;, \quad
  \cK_2^{(0)} = \Ga_2 - \cK_1^{(1)} \Ga_0 - \cK_0^{(1)} \Ga_1
    - \cK_0^{(2)} \Ga_0^2 \;.
\end{equation}
These expressions should be consistent with the $\om$-pole hypothesis so that
higher order poles in $\om$, possibly occurring in the r.h.s.\ of
eq.~(\ref{gammaPolePart}), should cancel out.

Furthermore, by the $\om$-pole hypotesis, we determine the $\om$-pole part of
$\cK_{gg,1}$ by a subtraction of the BFKL kernel, as follows:
\begin{equation}
  \label{omegaPolePart}
  \big(\zero\cK_1\big)_{gg}
  = (2\CA)^2 K_1^\BFKL - \big(\zero\cK_0 \, \one\cK_0\big)_{gg} \;.
\end{equation}
Once again, this should be consistent with the $\ga$-pole hypothesis, so that
quadratic (and possibly cubic) $\ga$-poles in $K_1^\BFKL$ should cancel out on
the r.h.s., and furthermore the simple pole should be consistent with
eq.~(\ref{gammaPolePart}), that is
\begin{equation}
  \big(\zero\cK_1^{(0)}\big)_{gg}
  = (2\CA)^2 K_1^{\BFKL\,(0)} - \big(\zero\cK_0 \, \one\cK_0\big)^{(0)}_{gg}
  = (\zero\Ga_1)_{gg} - \zero\big( \cK_0^{(1)} \Ga_0 \big)_{gg} \;.
  \label{simplePoleBFKL}
\end{equation}

Let us start proving the consistency relation for
eq.~(\ref{gammaPolePart}). Generally speaking, they are equivalent to recursive
relations on the $\om$-singuarities of $\Ga_n$ or, in other words, to the
resummation formulas proved in sec.~\ref{ss:rf}.
For instance, the assumed absence of $\om$-poles in $\big(\cK_n^{(0)}\big)_{qa}$
implies the NL$x$ resummation formulas for $\big(\Ga_n\big)_{qa}$:
\begin{align}
 \big(\Ga_1\big)_{qa} &\simeq \big(\cK_0^{(1)}\big)_{qg} \;
 \zero\big(\Ga_0\big)_{ga} \;, \qquad (\text{at } \frac{\as^2}{\om} \text{ level})
  \label{Gamma1qa} \\
 \big(\Ga_2\big)_{qa} &\simeq \big(\cK_0^{(2)}\big)_{qg} \;
 \zero\big(\Ga_0^2\big)_{ga} \;, \qquad (\text{at } \frac{\as^3}{\om^2} \text{ level})
  \label{Gamma2qa}
\end{align}
as predicted by eqs.~(\ref{gaqgNL}, \ref{gaqqNL}).

The $gq$, $gg$ entries are slightly more complicated. At order $\as^2/\om$,
$\zero\cK_{gq,1}^{(0)}$ is determined by eq.~(\ref{gammaPolePart}), so that no
consistency condition arises in the $gq$ entry. However, $\zero\cK_{gg,1}^{(0)}$
is already determined by eq.~(\ref{omegaPolePart}), so that the consistency
condition (\ref{simplePoleBFKL}) arises. The latter is verified because the
simple-pole part of $K_1^\BFKL$ is simply $\ga_{+,1}$~\cite{CaCi98}, so that
eq.~(\ref{simplePoleBFKL}) reduces to the identity (\ref{eq:ggConsCons}), which
implies, at order $\as^2/\om$,
\begin{align}
 \ga_{+,1} - \Ga_{gg,1} &= \big( \zero\cK_0^{(1)} \, \one\cK_0^{(0)}
 + \zero\cK_0^{(0)} \, \one\cK_0^{(1)} \big)_{gg} - \big( \zero\cK_0^{(1)} \,
 \one\cK_0^{(0)} + \one\cK_0^{(1)} \, \zero\cK_0^{(0)} \big)_{gg} \nonumber \\
 &= \left[ \zero\cK_0^{(0)} , \one\cK_0^{(1)} \right]_{gg}
 = \frac{2\CF}{\om} \big(\one\cK_0^{(1)}\big)_{qg} \;,
 \label{idqg1}
\end{align}
where the r.h.s.\ reduces, by eq.~(\ref{Gamma1qa}), to
$\frac{\CF}{\CA}\Ga_{qg,1}$, as it should.

Furthermore, at order $\as^3/\om^2$ we have consistency conditions for
$\Ga_{gg,2}$ and $\Ga_{gq,2}$. The former is identically satisfied, by some
algebra similar to eq.~(\ref{idqg1}), because
$\ga_{+,2}^\NL = \Ga_{gg,2}^\NL + \frac{\CF}{\CA}\Ga_{qg,2}^\NL$ as
given in eq.~(\ref{gammaPlus}). The latter is instead non trivial and, after a
similar algebra, reduces to
\begin{equation}
  \label{NNLOconstraint_bis}
   \Ga_{gq,2}^{\NL} = \frac{\CF}{\CA} \Ga_{gg,2}^{\NL} \;,
\end{equation}
as already proved in eq.~(\ref{eq:NNLOconstraint}), with the same consequences.

We finally note that cubic and quadratic $\ga$-poles are absent in
(\ref{omegaPolePart}) because of the identity, valid up to order $1/\ga^2$,
\begin{align}
 (2\CA)^2 K_1^\BFKL &= (2\CA)^2\left(\chi_1-\chi^\run-\half\chi_0\chi_0'\right)
 \nonumber \\
 &\simeq \big(\zero\cK_0^{(0)}\big)_{gg} \big(\one\cK_0^{(0)}\big)_{gg}
 + \big(\zero\cK_0^{(0)}\big)_{gq} \big(\one\cK_0^{(0)}\big)_{qg} \;.
\end{align}
Here the cubic poles at $\ga=0$ already cancel out in the l.h.s., because of the
$\half\chi_0\chi_0'$ subtraction needed to switch energy-scale $k k_0 \to k^2$,
due to the $\om$-shift~(\ref{eigenvChange}). The remaining quadratic poles are
given by the r.h.s., because of normal collinear factorization, and of absence
of $1/\ga$ singularities in $\dot{\chi}_0$ at energy-scale $k^2$
(eq.~(\ref{chi0dot})).

\section{Characteristic features of the resummed Green's function\label{s:cfrgf}}

In this section we present numerical results of some phenomenologically relevant
quantities which can be obtained by using the matrix kernel
$\cK = \alh \cK_0 + \alh^2 \cK_1$ developed in the previous sections. We recall
that the final form of $\cK_0$ and $\cK_1$ can be found in
eqs.~(\ref{eq:chi_mat_0_all_prerequisites}) and (\ref{eq:K1ansatz})
respectively, and that the detailed implementation of the running coupling is
found in eq.~(\ref{runningKernel}).

We state once more that our matrix kernel incorporates exactly the DGLAP and
BFKL properties at NLO and NL$x$ accuracy. However, in order to see the impact
of the NLO contributions and to compare with previous resummation approaches, we
will consider also results obtained from the kernel with only LO anomalous
dimensions  (but still in NL$x$ approximation). The corresponding kernel ---
which we refer to as NL$x$-LO model --- is built with the same $\cK_0$ given in
eq.~(\ref{eq:chi_mat_0_all_prerequisites}) but with $\cK_1$ including only the
$\tilde{K}_1$ term in the $gg$ entry, as can be read from
eq.~(\ref{eq:our_K1_struct}) by setting $\widetilde{\Ga}_1 = 0$.

\subsection{Hard Pomeron exponent}

We shall first investigate the high-energy $s \to +\infty$ behaviour of the
$A+B \to X$ differential cross section given in eq.~(\ref{heFact}) at fixed and
equal value of the two hard scales $k^2 \simeq k_0^2$, by determining the growth
exponent (hard Pomeron) $\oms$ in the limit of frozen coupling. In this limit,
we can use the representation~(\ref{eq:resolvent}) for the Green's
function $\G(\kt,\kt';\om)$ and, by using the spectral decomposition introduced
in sec.~\ref{ss:adm}, we obtain (the $\as$-dependence is understood)
\begin{equation}\label{xSecEq}
 \frac{\dif\sigma(k \simeq k_0)}{\dif^2\kt \, \dif^2\kt_0} = \sum_{l\in\{+,-\}}
 \int \frac{\dif\ga}{2\pi\ui} \;
 \int \frac{\dif\om}{2\pi\ui} \left(\frac{s}{k k_0}\right)^\om
 \frac{h^A(\kt,\om) \, \Pi_l\big(\ga,\om\big) \, h^B(\kt_0,\om)}
      {1-\eta_l\big(\ga,\om\big)} \;.
\end{equation}
The $\om$-integral gets contributions from the singularities (labelled by the
index $m$) of the integrand at $\om = \bar{\om}_{l,m}(\ga)$ due to the
vanishing of the denominator
\begin{equation}\label{vanishDen}
  1-\eta_l\big(\ga,\bar\om_{l,m}(\ga)\big) = 0 \;,
\end{equation}
thus providing
\begin{equation}\label{xSecEq1int}
 \frac{\dif\sigma(k \simeq k_0)}{\dif^2\kt \, \dif^2\kt_0} = \sum_{l\in\{+,-\}}
 \sum_{m\in M_l} \int \frac{\dif\ga}{2\pi\ui} \;
 \left(\frac{s}{k k_0}\right)^{\bar{\om}_{l,m}(\ga)}
 \frac{h^A(\kt,\bar\om) \, \Pi_l\big(\ga,\bar\om\big) \, h^B(\kt_0,\bar\om)}
      {-\partial_\om\eta_l\big(\ga,\bar\om\big)} \;.
\end{equation}
In the limit $s \gg k k_0$ the $\ga$-integral is dominated by the saddle-point
$\ga = \gas$ such that
\begin{equation}\label{gammaSaddlePoint}
  \frac{\dif}{\dif\ga} \bar{\om}_{l,m}(\gas) = 0
\end{equation}
for the particular values of $l$ and $m$ such that $\bar{\om}_{l,m}(\gas)$ is
maximum. It turns out that those values correspond to the leading-twist
component 
of the ($l=+$)-branch of the eigenvalue function $\eta_+$,
namely the solutions of $\eta_+(\gas,\oms) = 1$ with $\gas \to 0$ for $\as \to 0$.
As a result, in the high-energy limit the cross section has the power-like behaviour
\begin{equation}\label{sumOfPowers}
  \frac{\dif\sigma(k \simeq k_0)}{\dif^2\kt \, \dif^2\kt_0}
  = C_\mathrm{s} \left(\frac{s}{k k_0}\right)^{\oms}
\end{equation}
where the process dependent coefficient $C_\mathrm{s}$ is constant or at most
logarithmic in $s$, and the growth exponent $\oms$ is determined by the
conditions (\ref{vanishDen},\ref{gammaSaddlePoint}).

We can recast eq.~(\ref{gammaSaddlePoint}) into an equivalent relation for the
function $\partial_\ga\eta_l$. In fact, by taking the total $\ga$-derivative of
eq.~(\ref{vanishDen}) we can express
\begin{equation}\label{derOmBar}
 \frac{\dif \bar{\om}}{\dif\ga}
  = - \frac{\partial_\ga\eta}{\partial_\om\eta}
\end{equation}
thus obtaining the following conditions for the hard Pomeron exponent:
\begin{subequations}\label{omsCondition}
\begin{align}
    \quad\eta_+(\gas,\oms) &= 1 \\
    \partial_\ga\eta_+(\gas,\oms) &= 0 \;.
\end{align}
\end{subequations}
The above conditions in turn can be translated into analogous conditions for the
determinant of the operator $1-\cK$. In fact, from the relations
\begin{subequations}
\begin{align}
 \det(1-\cK) &= (1-\eta_+)(1-\eta_-) \label{factDet} \\
 \partial_\ga \det(1-\cK) &= - [ (1-\eta_+)\partial_\ga\eta_- \, + \,
 (1-\eta_-)\partial_\ga\eta_+ ]
\end{align}
\end{subequations}
eqs.~(\ref{omsCondition}) are equivalent to
\begin{subequations}\label{detCondition}
\begin{align}
   \quad\,\det[1-\cK(\gas,\bar{\om}_i)] &= 0 \\
   \partial_\ga\det[1-\cK(\gas,\bar{\om}_i)] &= 0 \;.
\end{align}
\end{subequations}

\begin{figure}[htbp!]
\centering{
  \includegraphics[width=0.7\textwidth]{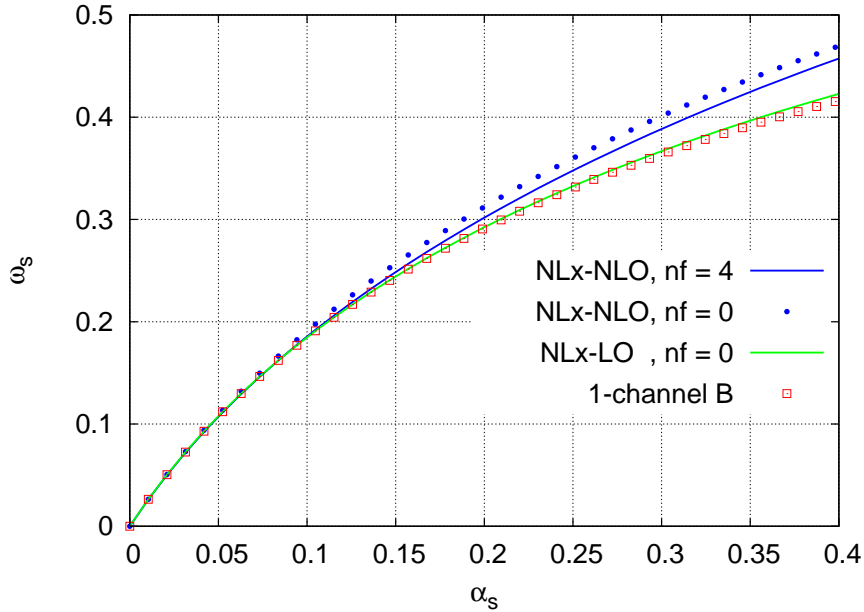}\\}
  \caption{\it Hard Pomeron exponent $\oms$ obtained in the NL$x$-NLO matrix formulation
    with $\nf = 4$ (solid blue) and $\nf = 0$ (dotted blue). The one-channel
    results~\cite{rgiggf} are also shown (red squares) and compared to those
    of the matrix model in NL$x$-LO approximation with
    $\nf = 0$ (solid green). The calculation is done in the fixed coupling case.
    \label{f:oms}}
\end{figure}

We have numerically solved the implicit equations (\ref{detCondition}) in our
matrix formulation in a range of $\as$ up to 0.4, both at NL$x$-NLO and NL$x$-LO
accuracy, for two values of $\nf = 0$ and 4. The results for $\oms$ versus $\as$
are shown in fig.~\ref{f:oms}, where we compare with results obtained from our
previous one-channel approach.

The NL$x$-LO curve at $\nf = 0$ almost overlaps to the old one-channel result,
thus showing the stability of the matrix formulation and the continuity with
the one-channel approach. In fact, since at $\nf = 0$ the kernel is diagonal, only
the $gg$ entry determines $\oms$. The small discrepancy is due to: {\it (i)} the
momentum-conserving factor in front of $\tilde{\chi}_1$ in
eq.~(\ref{eq:our_K1_struct}) (it was a plain $1/\om$ in the one-channel case);
{\it (ii)} a non vanishing two-loop anomalous dimension (only for the low-energy
part, actually) (we enforced vanishing anomalous dimension in the one-channel
case). By including the NLO contributions we obtain a moderate increase of the
Pomeron intercept, which slightly diminishes when quarks are also taken into
account.

\subsection{Effective characteristic function(s)}

As already noted in the previous section, the contributions to the integral
representation of the cross section stem from those values of $\om$ and $\ga$
such that
\begin{equation}\label{implEq}
  \det[1-\cK(\as,\ga,\om)] = 0 \;.
\end{equation}
which provides a relation between the moment index $\om$ and the anomalous
dimension variable $\ga$. Solving eq.~(\ref{implEq}) for either $\om$ or $\ga$
defines the {\em effective characteristic function} and its dual {\em effective
  anomalous dimension}
\begin{equation}\label{chiEff}
  \om = \chi_\eff(\as,\ga) \;, \qquad \ga = \ga_\eff(\as,\om) \;.
\end{equation}
While in the one-channel case we have only one perturbative branch of those
functions, corresponding to the BFKL eigenvalue function $\chi_+(\ga)$ and to
the larger eigenvalue $\ga_+(\om)$ of the anomalous dimension matrix, in the
matrix formulation we expect two branches. The second branch corresponds to the
smaller eigenvalue $\ga_-(\om)$ which is dual to a second effective
characteristic function $\chi_-(\ga)$.

\begin{figure}[htp!]
\centering{
  \includegraphics[width=0.7\textwidth]{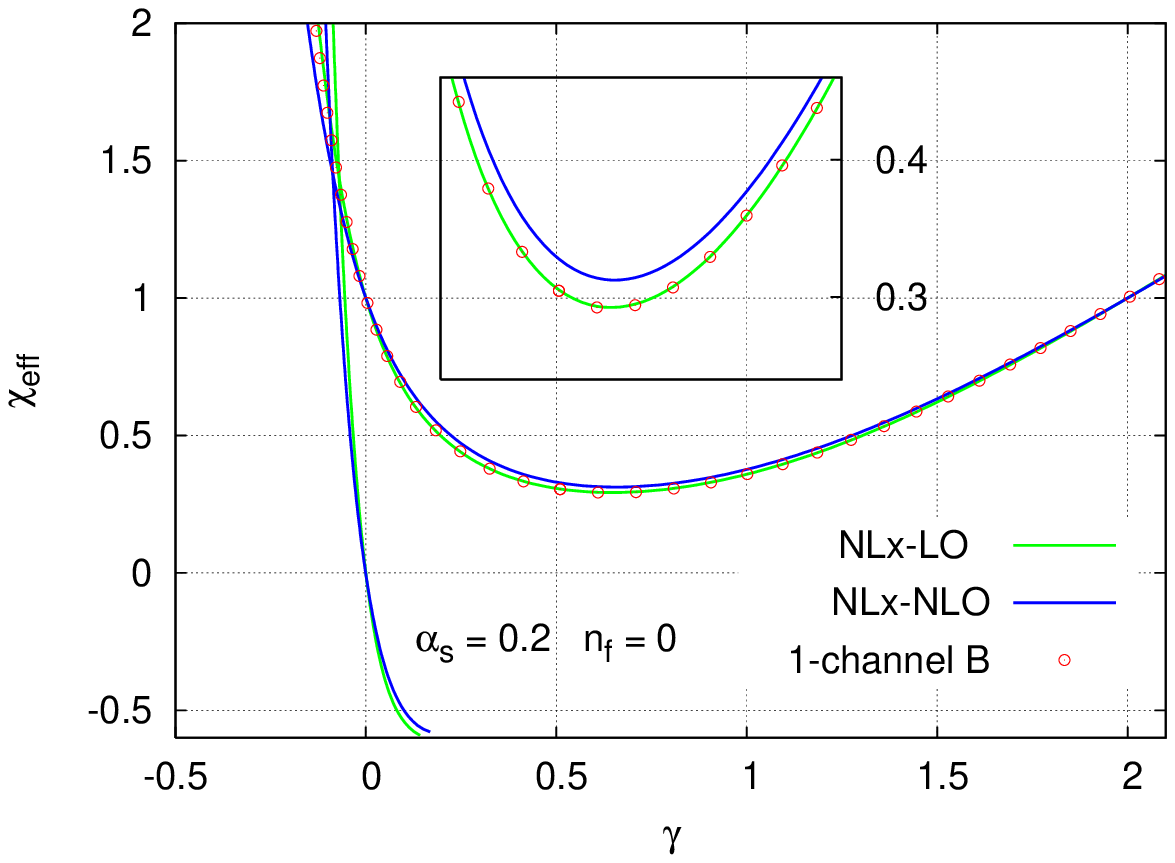}\\
  {\it a)}\\
  \includegraphics[width=0.7\textwidth]{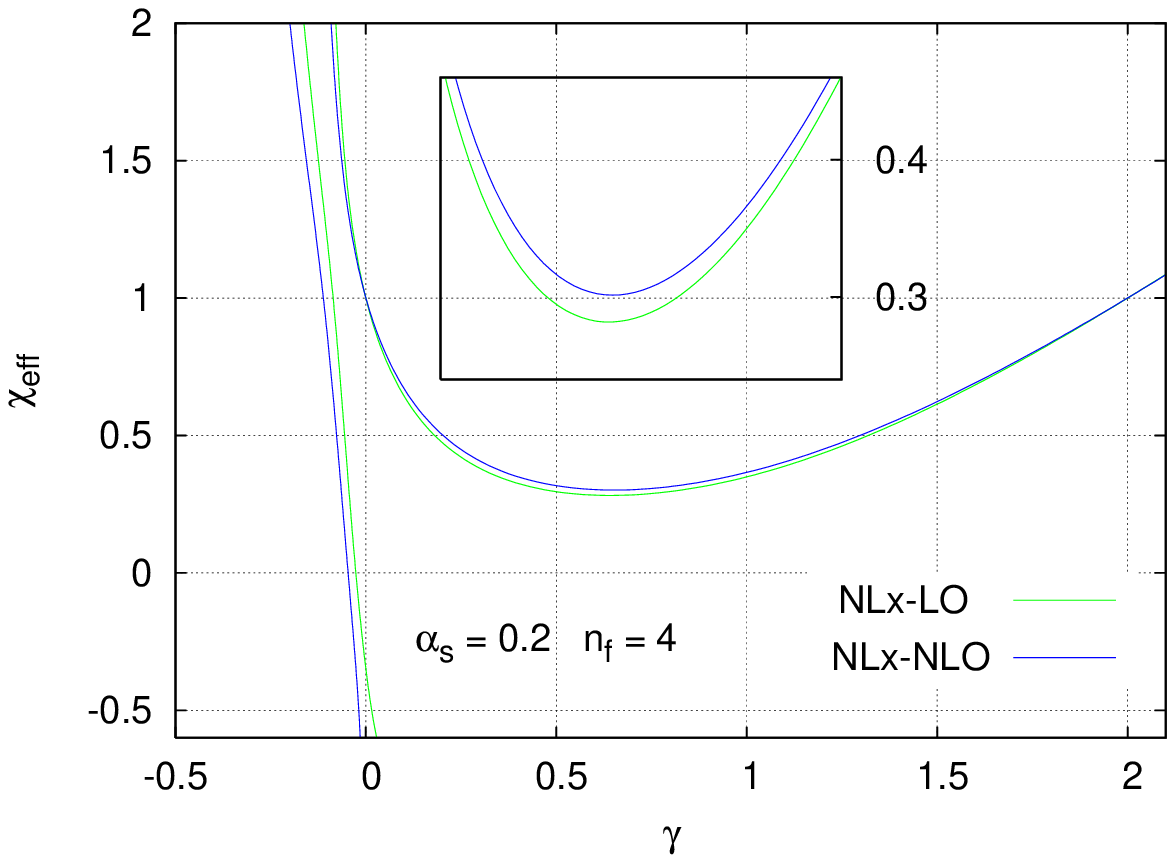}\\
  {\it b)}\\
}
\caption{\it Effective eigenvalue function obtained in the matrix formulation with
  full NL$x$-NLO accuracy (blue) and with NL$x$-LO accuracy (green) for $\nf=0$
  (a) and $\nf=4$ (b). The steeply decreasing curves on the left side of
  each plot represent the minus-branches $\chi_-$, while the curves with a
  minimum around $\ga \simeq 0.5$ represent the plus-branches $\chi_+$. The red
  circles reproduce the one-channel result in scheme B of ref.~\cite{rgiggf}. The
  calculation is done at fixed coupling $\as=0.2$.
    \label{f:chiEff}}
\end{figure}

In fig.~\ref{f:chiEff} we show the two branches of the effective characteristic
functions obtained in the NL$x$-NLO and NL$x$-LO cases. We have considered here
the asymmetric $\om$-shift corresponding to the energy-scale $k^2$, with
$\as=0.2$. The $\chi_+$'s are characterised by the typical minimum around $\ga
\simeq 0.5$ whose value is nothing but $\oms(\as)$. On the other hand, the
$\chi_-$'s appear as steeply decreasing functions located around $\ga \lesssim 0$ in
the region shown in our plots.

The continuity of the resummation procedure when going from the one-channel to
the two-channel formulation at $\nf=0$ is illustrated in fig.~\ref{f:chiEff}a by
the overlapping of the ($+$)-branch of the NL$x$-LO curve to the circles
corresponding to the one-channel scheme-B effective eigenvalue function.  The
NLO terms provide a slight increase of the ($+$)-branch in the region $0< \ga <2$,
and a small decrease of the ($-$)-branch at $\ga < 0$. At $\nf = 0$ there is a
crossing point of the two branches at negative $\ga$ in either approximations.

The inclusion of quarks removes the crossing (with a mechanism similar to the
degenerate level splitting in quantum mechanics) causing $\chi_-$ to be always
on the left of $\chi_+$, as can be seen in fig.~\ref{f:chiEff}b. Quantitatively,
the quark contribution lowers both $\chi_+$ in the region around the minimum (compare
the two inserts in fig.~\ref{f:chiEff}) and $\chi_-$.

Note the two fixed points at $(\ga,\om=\chi_\eff)$ = $(0,1)$ and $(2,1)$ of the
($+$)-branches. In the one-channel case these fixed points corresponds to momentum
conservation in the collinear and anti-collinear limits respectively. In the
two-channel formulation they imply that the anomalous dimension eigenvalue
$\ga_+(\om=1) = 0$; however, momentum sum rule is satisfied provided the
corresponding left-eigenvector $\bar{v}_+ = \row{r}{1}$ of the anomalous
dimension matrix (cf.~eq.~(\ref{vplus})) be $\row{1}{1}$ at $\om=1$, i.e.,
provided $r(\om=1) = 1$.

Actually, our matrix model presents a small violation of the momentum sum rule.
In fact, by exploiting the fact that $\ga_+(\om=1) = 0$ and by using
eqs.~(\ref{eq:DGLAPeq}) and (\ref{GammaSpectral}) we have (at $\om=1$)
\begin{equation}\label{momViol}
 \dot{q} + \dot{g} = \row{1}{1} \Ga \column{q}{g} =
 \ga_- \, \frac{1-r}{1+r\rho} (q-\rho g) \;.
\end{equation}
The computation of the prefactor $\ga_- \, (1-r) \rho/(1+r\rho)$ versus $\as$
shown in tab.~\ref{t:mnc} gives us an estimate of the relative amount of
momentum non-conservation; the violation is of order $\as^2$ for the NL$x$-LO
scheme, and of order $\as^3$ for the scheme with NLO terms included.

\begin{table}[htp!]
  \centering
  \begin{tabular}{|c|c|c|c|c|} 
\hline
$\as$   &   NL$x$-LO  &   NL$x$-NLO   & NL$x$-LO/$\as^2$ & NL$x$-NLO/$\as^3$ \\
\hline
0.025   &   0.00019   &   0.0000031   &   0.302       & 0.199          \\
0.050   &   0.00072   &   0.0000208   &   0.287       & 0.167          \\
0.100   &   0.00260   &   0.0001303   &   0.260       & 0.130          \\
0.150   &   0.00534   &   0.0003437   &   0.237       & 0.102          \\
0.200   &   0.00872   &   0.0006107   &   0.218       & 0.076          \\
\hline
  \end{tabular}
  \caption{\it Estimate of momentum sum rule violation. The quantity
    $\ga_- \, (1-r)\rho/(1+r\rho)$ at $\om = 1$ has been computed for various values
    of $\as$ (column 1) in the NL$x$-LO (column 2) and NL$x$-NLO (column 3)
    schemes. Column 4 (resp.\ 5) shows that the NL$x$-LO (NL$x$-NLO) violation
    is of order $\as^2$ ($\as^3$). \label{t:mnc}}
\end{table}

\section{Numerical results with running coupling\label{s:nrrc}}

In this section we shall present results obtained by solving
eq.~(\ref{eq:greenFunc}) in $(x,\kt)$-space, including a running coupling. The
basic structure of the ensuing integral equation follows from
eq.~(\ref{runningKernel}) and reads ($Y \equiv \log 1/x$)
\begin{align}
 \G_{ab}(Y;k,k_0) &= \delta_{ab} \Theta(Y)\frac{\delta(k^2-k_0^2)}{\pi} + \sum_c
 \int_x^1 \frac{\dif z}{z} \int \dif k'{}^2 \Big[\alh(q^2)
   \delta_{ag}\delta_{gc} K_0(z;k,k')
 \nonumber \\ \label{integralEq}
 &\qquad + \alh(k_>^2)
   \cK_{\mathrm{coll},ac}(z;k,k') + \alh^2(k_>^2)\cK_{1,ac}(z;k,k') \Big]
 \G_{cb}\big(\log\frac{z}{x};k',k_0\big)
\end{align}
(the explicit expressions of the kernels $K_0$, $\cK_{\mathrm{coll}}$ and
$\cK_1$ in the equation above are given in app.~\ref{a:kcf}).  We shall extract
Green functions and splitting functions, using the methods described
in~\cite{rgiggf,extending,CCSFact}. In both the coupling and the kernels we use
a fixed number of flavours, $n_f=4$.  The coupling runs with a 2-loop $\beta$
function, and is normalised such that $\as(3\GeV) = 0.256$. The infrared region
of the coupling is regularized by setting it to zero for scales
$\mu < \mu_0 = 0.75 \GeV$.

The results that we shall show are those of the model described above
(NL$x$-NLO), and also those for a model in which the higher twist part
of $K_{0,qg}$ has been supplemented with (symmetric) $1/(1+\ga)^2$ and
$1/(2+\om-\ga)^2$ terms
\footnote{More precisely, the higher-twist kernel reads
$\chiht^\om(\ga) = \frac{134}{81}(1+\ga)^{-1} - \frac{32}{27}(1+\ga)^{-2}
+ [(1+\ga) \to (2+\om-\ga)]$.}
so that not only the $\as^2/\om$ but also
$\as^3/\om^2$ terms of the $qq$, $qg$ and $gg$ splitting functions are
in the $\MSbar$ scheme.%
\footnote{The $gq$ term is almost in the
  $\MSbar$ scheme, the only difference being a small $N_c$-suppressed 
  contribution of relative order $n_f/N_c^2$, corresponding to the violation of
  eq.~(\ref{NNLOconstraint_bis}) in the NNLO $\MSbar$ splitting
  functions.}
We shall denote this second model NL$x$-NLO$^+$.  We shall also compare
to results obtained in our earlier single-channel
work~\cite{extending} (scheme B), where we used a 1-loop coupling,
$n_f=0$ in the kernel (but $n_f=4$ in the coupling) and for which the
NLO piece of the effective $P_{gg}$ splitting function was identically
zero.

\subsection{Green functions}
\label{sec:green-fns}

\begin{figure}
  \centering
  \includegraphics[width=0.6\textwidth]{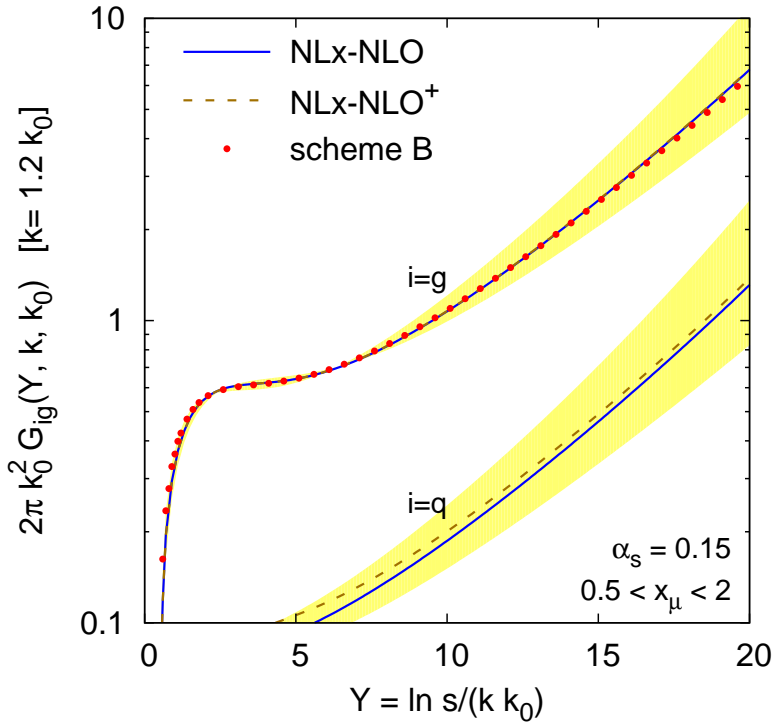}
  \caption{\it Gluon-induced part of the Green function for the NL$x$-NLO and
    NL$x$-NLO$^+$ models, compared to the results of~\cite{extending} (scheme B).
    For the models of this paper both $\G_{gg}$ and $\G_{qg}$ are shown.
    The value chosen for the coupling, $\as=0.15$, corresponds to
    $k_0\simeq 20 \GeV$. The band indicates the spread in the result for
    the NL$x$-NLO model when varying the renormalisation scale in the
    range $0.5 < x_\mu < 2$.}
  \label{fig:green}
\end{figure}

The Green function for the matrix evolution is itself a matrix in
flavour space. Physically the most interesting part is that involving
gluonic sources, and this is shown in fig.~\ref{fig:green}.

The old and new resummations give nearly identical results for the
$\G_{gg}$ part of the result, indicative of the stability of the
resummation procedure. Furthermore, the differences between them are
much smaller than renormalisation scale uncertainty, which grows with
$Y$.  The growth with $Y$ of the scale uncertainty can be understood
as an indication of underlying scale dependence of the
effective BFKL exponent.

The $\G_{qg}$ channel can be given only within the new
resummations. As would be expected, the quark component is suppressed
by a factor $\sim \as$ compared to the gluon component.  A
consequence of the fact that the quarks are only generated radiatively
is the scale dependence in their normalisation as well as in their
growth with $Y$.  We note that the change induced by the NNLO
scheme-dependent higher-twist part of $K_{0,qg}$ (NL$x$-NLO$^+$ versus
NL$x$-NLO) is small, despite the fact that the region $k\sim k_0$ that
we study is that most likely to be sensitive to this higher-twist
contribution.

\subsection{Splitting functions}
\label{sec:split-fns}

The extraction of splitting functions is carried much in the same way
as in the one-channel case described in~\cite{CCSFact,rgiggf}. There a
special (infrared) inhomogeneous term was included in the equation for
the Green function such as to ensure that the resulting integrated
gluon distribution satisfies $xg(x,\mu^2)=1$, independently of $x$,
for $\mu^2$ set equal to some given $Q^2$.  With that inhomogeneous
term fixed, the $xP_{gg}(x,Q^2)$ splitting function was then obtained
as $\frac{\partial}{\partial\ln x} \frac{\partial}{\partial\ln\mu^2} xg(x,\mu^2)|_{\mu^2 = Q^2}$.
In the matrix case, we have a 2-component vector of inhomogeneous
terms: we can choose it such that $xq(x,\mu^2)=0$, $xg(x,\mu^2)=1$ for
$\mu^2=Q^2$, in which case we obtain
\begin{equation}
  \label{eq:qqgg}
  \left(\!
    \begin{array}{c}
      xP_{qg}\\
      xP_{gg}
    \end{array}\!
  \right)
   = 
   \frac{\partial}{\partial\ln x} \frac{\partial}{\partial\ln\mu^2}
   \left.
     \left(\!
       \begin{array}{c}
         xq(x,\mu^2)\\
         xg(x,\mu^2)
       \end{array}\!
     \right)
   \right|_{\mu^2 = Q^2}\,.
\end{equation}
Alternatively we can set the inhomogeneous terms so as to ensure that
$q(x,\mu^2)=1$, $g(x,\mu^2)=0$ for $\mu^2=Q^2$ and we then extract
$P_{qq}$ and $P_{gq}$.

\begin{figure}
  \centering
  \includegraphics[width=\textwidth]{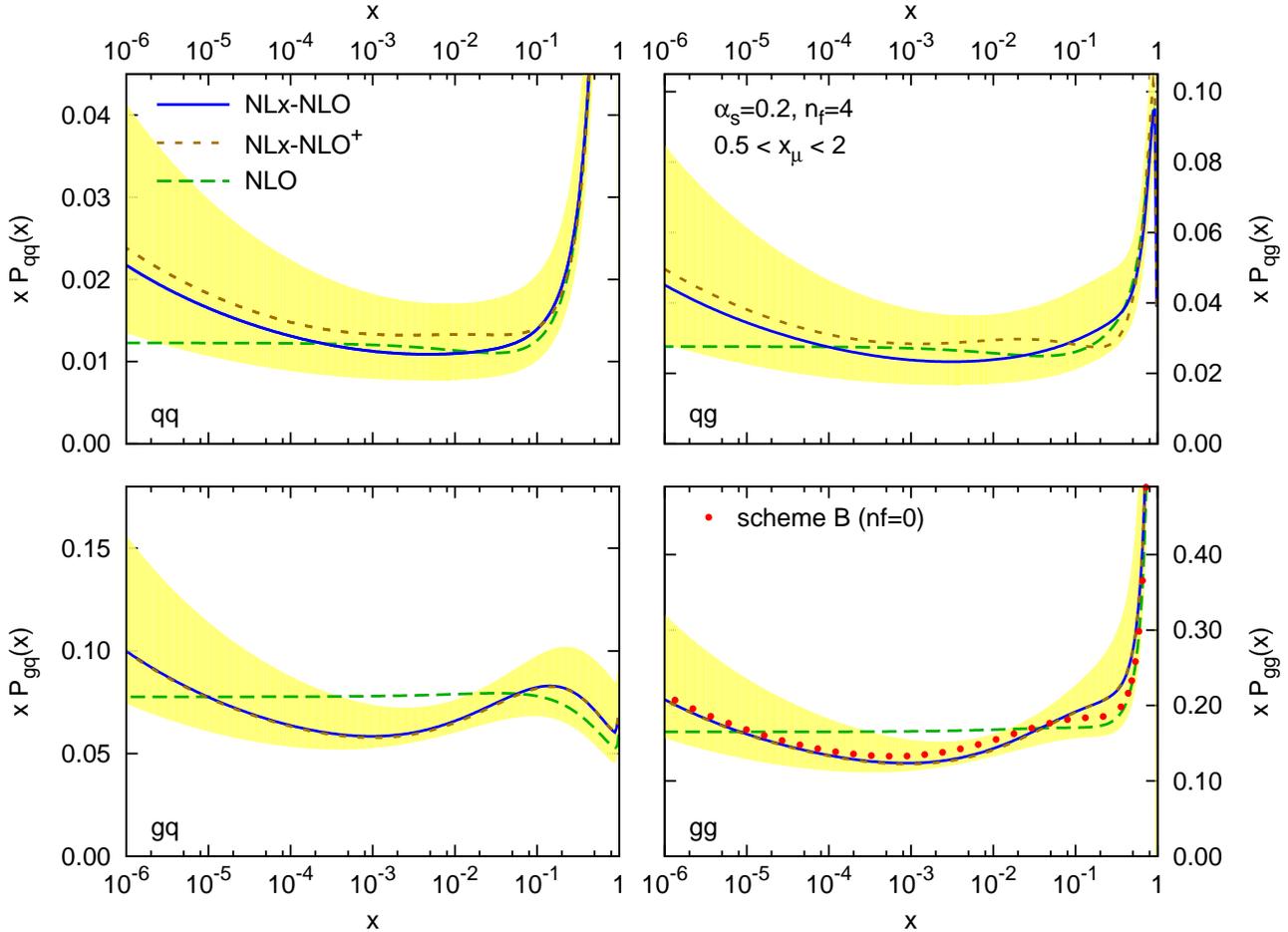}
  \caption{\it The matrix of NL$x$-NLO (and NL$x$-NLO$^+$) splitting functions
    together with their scale uncertainty and the NLO splitting
    functions for comparison. In the $gg$ channel, we also show the
    old scheme~B result ($n_f=0$, no NLO contributions, 1-loop
    coupling) of~\cite{rgiggf}. The band corresponds to the span of
    results (NL$x$-NLO) obtained if one chooses $x_\mu=0.5$ and
    $x_\mu=2.0$.}
  \label{fig:4tile-NLxNLO}
\end{figure}

The matrix of effective splitting functions as determined with this
method is shown in fig.~\ref{fig:4tile-NLxNLO}, for both our kernels 
and with a scale $Q\simeq 6 \GeV$, giving $\as(Q^2) = 0.2$. For reference we
plot also the exact NLO splitting functions and our previous results
for the single-channel evolution.  Considering first the $gg$ channel,
the results are rather similar to the old ones, and in particular
maintain the characteristic dip~\cite{gluonsf} around $x=10^{-3}$ that
has been seen also by the authors of~\cite{ABF_improved}. This dip
is present also in the $gq$ channel, and indeed the $gq$ channel is
rather similar in a range of features to the $gg$ channel, which is
natural since it is largely driven by the summation of the $g\to g$
branching. Among the common features is the slight but noticeable
difference compared to the NLO splitting functions at moderate $x$ ($x
\gtrsim 0.1$). The detailed origin of this characteristic is not
really understood, but may well be connected with the fact that the
various pieces of the NLO $gg$ splitting function are effectively
placed in different parts of our evolution kernel, and then subjected
to non-trivial (higher order) non-linear effects in their
recombination into the final effective splitting
function.\footnote{The greater similarity between the large-$x$ scheme
  B kernel and the NLO results is an artefact related to the different
  $n_f$ values used in the old scheme~B results and the new matrix
  evolution.}

Concerning the scale dependence of the $P_{gj}$ channels, as for the
Green function, it grows significantly towards small $x$, and again
this is a sign of scale dependence of the small-$x$ intercept. One
notes non-negligible scale dependence also at moderate $x$. Generally,
down to moderately small $x$, the scale dependence of the NL$x$-NLO
splitting functions (all channels) is rather similar to that (not
shown) of the plain NLO splitting functions.

The two $P_{qj}$ channels differ fundamentally from the $P_{gj}$
channels in that they are non-zero at small $x$ starting only at NL$x$
and at NLO. Thus there is a sense in which our NL$x$-NLO treatment is
effectively a leading order treatment for these channels, at least as
concerns their normalisation (the small-$x$ growth is driven by
iterations in the gluon channel, so one expects this to be under
better control). This is visible in the much larger scale dependence
for these channels. They also have some (modest) sensitivity to the
difference between the NL$x$-NLO and NL$x$-NLO$^+$ kernels, whereas in the
$P_{gj}$ channels there was almost no sensitivity to this difference
(even though the difference is NL$x$ in all channels).
A general feature of the $P_{qj}$ splitting functions is that they are
rather similar to the NLO splitting functions (more so than in the
gluon channel). In particular, though like the $P_{gj}$ splitting
functions they have a dip around $x=10^{-3}$, this dip is considerably
shallower. The conclusion here is that the NLO $P_{qj}$ splitting
functions can probably be considered a good approximation to the full
splitting functions for $x$ as low as $10^{-4}$.

An important cross-check of the methods used to extract the splitting
functions is that the results should be independent (modulo
higher-twist contributions) of the infrared regularisation of the
coupling, i.e.\ independent of the scale $\mu_0$ below which the
coupling is set to zero. To this end we have extracted the splitting
functions with $\mu_0$ increased from $0.75\GeV$ to $1\GeV$
(corresponding to reducing $\as(\mu_0)$ from $0.58$ to $0.45$) and
find that the results change only by a few percent.\footnote{One may
  also reduce $\mu_0$, however for $\mu_0 \lesssim 0.6\GeV$,
  $\as(\mu_0)$ then becomes so large that numerical instabilities
  develop, and it becomes impossible to extract meaningful results.}
As in previous work~\cite{CCSFact,rgiggf} we find that these
factorization violations scale roughly as $1/Q$ rather than as
$1/Q^2$, a characteristic perhaps attributable to resummation effects,
which could quite conceivably modify typical collinear $1/Q^2$
power-suppressed effects such that they become $1/Q^{2-2\ga}$ with an
effective $\ga\simeq1/2$.

\begin{table}
  \centering
  \begin{tabular}{c|c|r@{.}l|r@{.}l|r@{.}l|r@{.}l|}
    \multicolumn{2}{c|}{}
    & \multicolumn{4}{c|}{NL$x$-NLO} & \multicolumn{4}{c|}{NL$x$-NLO$^+$}\\[1pt]
    $\as$ & $Q$ [GeV]  
                 & \multicolumn{2}{c|}{$\sum_j \Ga_{jq}(1)$} 
                 & \multicolumn{2}{c|}{$\sum_j \Ga_{jg}(1)$}
                 & \multicolumn{2}{c|}{$\sum_j \Ga_{jq}(1)$}
                 & \multicolumn{2}{c|}{$\sum_j \Ga_{jg}(1)$} \\[3pt] \hline
    $0.20$& 6    &  \mbox{ } 0&0079    &     \mbox{ }   -0&0059   
                 &  \mbox{ } 0&0074    &     \mbox{ }   -0&0055  \\ 
    $0.15$& 20   &  \mbox{ } 0&0021    &     \mbox{ }   -0&0015   
                 &  \mbox{ } 0&0018    &     \mbox{ }   -0&0012   \\ 
    $0.10$& 220  &  \mbox{ } 0&00012   &     \mbox{ }   -0&00003  
                 &  \mbox{ } 0&00006   &     \mbox{ }    0&00002   \\ 
  \end{tabular}
  \caption{\it Momentum sum-rule violation in the NL$x$-NLO and NL$x$-NLO$^+$
    models for three values of $\as$. The numerical uncertainty is
    roughly $\pm 1$ on the last digit of each result. }
  \label{tab:mom-sum-viol-running}
\end{table}

We close this section by showing in
table~\ref{tab:mom-sum-viol-running} the degree of momentum sum-rule
(MSR) violation in the splitting functions for three values of $\as$.
From just a small number of values it is difficult to
establish the exact scaling law,\footnote{Limits on the available
  numerical accuracy make it difficult to obtain reliable estimates of
  the MSR violations for smaller values of $\as$, because as $\as$
  decreases one needs ever higher relative accuracy to accurately
  determine the rapidly vanishing MSR-violating component.}  and in
particular it is difficult to determine the relative admixture of
higher-twist and perturbative components in the MSR violations.
Nevertheless, one sees that the MSR violation vanishes very rapidly as
$\as$ decrease, suggesting that a significant component of it is
non-perturbative in origin. This conclusion is borne out by studies
which show that the amount of MSR violation depends somewhat also on
$\mu_0$, the infrared cutoff scale for the coupling.

\section{Discussion}

We have proposed here a matrix evolution equation for the flavour singlet,
unintegrated quark and gluon densities, which generalizes the DGLAP and BFKL
equations in the relevant limits.

The matrix approach (secs.~\ref{sec:bmf} and \ref{sec:knla}) is supposed to unify collinear and high-energy
factorizations in both partonic channels, and is not necessarily guaranteed to
actually work, because of the various crossed consequences that the above
factorizations have: consider, for instance, the anomalous dimension
resummation formulae arising from $\kt$-factorization~\cite{QQvert,CaHa94} and
the $\ga\leftrightarrow 1+\om-\ga$ symmetry of the BFKL
kernel~\cite{Salam98,omExp} arising from collinear factorization. It is
therefore a nontrivial result of this paper that our resummed splitting
functions do satisfy collinear factorization in matrix form, as shown in
secs.~\ref{s:fcad} and \ref{s:nrrc}. In this respect, our approach defines, by the matrix evolution,
some unintegrated densities that are appropriate both in the collinear and in
the small-$x$ limits. It would be interesting to explore the relationship of
such explicit construction with alternative studies~\cite{Ciaf88,unintegrated}.

Furthermore, we want to incorporate exact low-order anomalous dimensions in
our matrix kernel, say in the $\MSbar$ scheme. We find, in this context, a new
kind of consistency relations on the kernels, due to a possible clash of exact
low-order expressions with a novel NL$x$ resummation formula for $\Ga_{gq}$,
arising in the matrix evolution (sec.~\ref{s:fcad}). We prove such relations to be
satisfied by our construction in the $\MSbar$ scheme at NLO, but marginally
violated by $n_f/N_c^2$-suppressed terms at NNLO. We are thus able to complete
our construction with exact NLO anomalous dimensions and NL$x$ kernel, and we postpone the
analysis of the NNLO accuracy, which is however nearly incorporated (in
the NL$x$ approximation) in our
NLO$^+$ version.

The frozen-$\as$ features of our matrix model are characterized by the
previously mentioned resummation formulae of sec.~\ref{s:fcad}, and by the hard Pomeron
exponent and effective eigenvalue functions of sec.~\ref{s:cfrgf}. One should notice the
basic continuity of our matrix approach with the single-channel case in the
$n_f=0$ limit, and the corresponding agreement of the leading effective
eigenvalue function with the ABF approach. Additionally, we provide here the
subleading effective eigenvalue at $n_f=4$, corresponding to the $\ga_-$
eigenvalue of the anomalous dimension.

We are finally able to provide the whole matrix of resummed splitting
functions in sec.~\ref{s:nrrc}. Roughly speaking, the outcome shows that resummation
effects are small in the $P_{qa}$ entries up to $x$-values as small as
$10^{-4}$, while the shallow dip is the main qualitative feature of both
$P_{ga}$ entries, with resummation effects starting below $x \simeq 10^{-3}$.

We notice that the above results are in the $\MSbar$ scheme up to NLO
(including approximately NNLO in their NLO$^+$ version), but generally differ
from it at higher orders. However, resummation effects in the scheme change
have been studied in~\cite{CCSS_MSbar} and turn out to
be of the order of the scale uncertainty. For this reason we believe that the
above results can be safely used in the study of structure functions and other
cross-sections, by supplementing them with the corresponding coefficient
functions or impact factors.

\section*{Acknowledgments}
We wish to thank Guido Altarelli, Richard Ball, Stefano Catani, Stefano Forte
and Al Mueller for various conversations on topics related to this paper.
We are grateful to the Galileo Galilei Institute for Theoretical Physics in
Arcetri for hospitality during the workshop on ``High Density QCD'', while
part of this work was being done.  This paper is 
supported in part by a PRIN grant from MIUR (Italy) and by the French
Agence Nationale de la Recherche, grant ANR-05-JCJC-0046-01. A.M.S  has been supported by the Polish Committee for Scientific Research
grant No. KBN 1 P03B 028 28.


\appendix

\section{Recursive expressions for the anomalous dimensions\label{a:read}}
In this appendix we show how eqs.~(\ref{eq:Gamma_from_cK}) are obtained.
Let us first rewrite eqs.~(\ref{eq:integratedDensities},\ref{eq:DGLAPeq}) in
$\ga$-space, by noting that $\ga = \partial_{\log Q^2}$:
\begin{equation}\label{DGLAPgammaSpace}
 \ga\, f_i(\ga,\om) = \ugd_i(\ga,\om)
 = \Ga_{ij}(\om) f_j(\ga,\om) + \text{h.t.}
\end{equation}
where ``h.t.'' stands for higher-twist contributions characterised by being
regular at $\ga = 0$. It follows that, in matrix notation,
\begin{equation}\label{gammaSqF}
 \ga^2 f = \ga (\Ga f + \text{h.t.}) = \Ga(\ga f) + \text{h.t.}
 = \Ga^2 f + \text{h.t.}
\end{equation}
and, by induction,
\begin{equation}\label{gammaPowF}
 \ga^n f = \Ga^n f + \text{h.t.} \;,
\end{equation}
Secondly, we consider eq.~(\ref{eq:genBFKLeq}) and expand the matrix kernel
$\cK$ in powers of gamma (according to the notations following
eq.~(\ref{eq:Kexpansion})), obtaining 
\begin{equation}\label{eqBFKLgammaExpnsn}
 \ugd = \left(\sum_{m=0}\cK^{(m)} \ga^{m-1}\right) \ugd + \ugd^{\mathrm{source}}
 = \sum_{m=0}\cK^{(m)} \ga^m f + \ugd^{\mathrm{source}}
 = \sum_{m=0}\cK^{(m)} \Ga^m f + \text{h.t.} \;,
\end{equation}
By comparing eqs.~(\ref{DGLAPgammaSpace}) and (\ref{eqBFKLgammaExpnsn}) we
derive the implicit equation
\begin{equation}\label{GammaKrel}
 \Ga = \sum_{m=0} \cK^{(m)} \Ga^m \;,
\end{equation}
which allows us to determine the effective anomalous dimension matrix $\Ga$
in terms of the matrix kernel $\cK$.

It is now straightforward to compute the perturbative coefficients $\Ga_n$
defined in eq.~(\ref{eq:GammaExpansion}). By expanding eq.~(\ref{GammaKrel}) to
first order in $\alh$ yields ($m=0$)
\begin{equation}
  \alh \Ga_0 = \alh \cK_0^{(0)}
\end{equation}\label{orderAlh1}%
from which eq.~(\ref{Gamma_0}) follows. By expanding eq.~(\ref{GammaKrel}) to
second order in $\alh$ yields ($m \leq 1$)
\begin{equation}
 \alh \Ga_0 + \alh^2 \Ga_1 = \alh \cK_0^{(0)} + \alh^2 \cK_1^{(0)}
 + \alh \cK_0^{(1)} \alh \Ga_0 \;.
\end{equation}
At frozen coupling, the operators $ \cK_0^{(1)}$ and $\alh$ commute. By then
collecting the $\order{\alh^2}$ terms and remembering that
$\Ga_0 = \cK_0^{(0)}$ we get
\begin{equation}\label{orderAlh2}
  \alh^2 \Ga_1 = \alh^2 ( \cK_1^{(0)} + \cK_0^{(1)} \cK_0^{(0)} )
\end{equation}
from which eq.~(\ref{Gamma_1}) follows. A similar iteration procedure produces
eq.~(\ref{Gamma_2}) and higher orders.

In the running coupling case, we have an additional commutator term starting
at second order, namely
\begin{equation}\label{commuTerm}
 \alh [ \cK_0^{(1)}, \alh] \cK_0^{(0)} \;.
\end{equation}
By using the expansion
\begin{equation}\label{alhExpnsn}
 \alh = \alh_\mu - \beta_0 \alh_\mu^2 \log\textstyle{\frac{k^2}{\mu^2}} + \order{\alh_\mu^3} \;,
\end{equation}
the commutator reads
\begin{equation}\label{commutator}
 [ \cK_0^{(1)}, \alh] = - \beta_0 \alh^2 [ \cK_0^{(1)}, \log\textstyle{\frac{k^2}{\mu^2}}]
 + \order{\alh^3} \;,
\end{equation}
thus producing, by eq.~(\ref{commuTerm}), a contribution of
order $\alh^3$ to the anomalous dimension matrix. Note however that the NL$x$
term of order $\as^3/\om^2$ vanishes, because the $gg$ entry of
 $\cK_0^{(1)}$  has no leading $1/\om$ term.

Extending the above procedure to higher orders, we see that at each order (say
N$^{n}$LO) the anomalous dimension gets a new term $\cK_n^{(0)}$ but also
a series of other terms which are combinations of the anomalous dimensions at
the lower orders $<n$.

\section{Splitting functions and anomalous dimensions}

Singlet anomalous dimensions and splitting functions at lowest order appear in
many of our formulas. The former are given by ($\Tf \equiv \Tr\nf$)
\begin{align}
  P_{qq,0}(z) &= C_F \left[\frac2{(1-z)_+} - 1 - z + \frac32 \delta(1-z) \right]
 \label{Pqq} \\
  P_{qg,0}(z) &= 2 \Tf [z^2 + (1-z)^2]
 \label{Pqg} \\
  P_{gq,0}(z) &= C_F \frac{1+(1-z)^2}{z}
 \label{Pgq} \\
  P_{gg,0}(z) &= 2\CA \left[ \frac1{z} + \frac1{(1-z)_+} -2+z-z^2 \right]
   + \frac{11\CA - 4\Tf}{6} \delta(1-z) \;.
 \label{Pgg}
\end{align}
The anomalous dimensions, i.e., the Mellin transforms of the splitting functions,
are given by
\begin{align}
  \Ga_{qq,0}(\om) &= \CF \left[ 2\psi(1)-2\psi(\om+1) - \frac1{\om+1}
  -\frac1{\om+2} + \frac{3}{2} \right]
 \label{gamma0_qq} \\
  \Ga_{qg,0}(\om) &= 2 \Tf \left( \frac1{\om+1} - \frac2{\om+2}
   + \frac2{\om+3} \right)
 \label{gamma0_qg} \\
  \Ga_{gq,0}(\om) &= \CF \left[ \frac2{\om}
   - \frac{2}{\om+1} + \frac1{\om+2} \right]
 \label{gamma0_gq} \\
  \Ga_{gg,0}(\om) &= 2\CA \left[ \frac1{\om} + \psi(1) - \psi(\om+1)
   - \frac2{\om+1} + \frac1{\om+2} - \frac1{\om+3} + \frac{11}{12} \right]
   - \frac{2\Tf}{3} \;.
 \label{gamma0_gg}
\end{align}
The gluon anomalous dimensions $\Ga_{gq,0}$ and $\Ga_{gg,0}$ are singular at
$\om = 0$. The regular parts are defined by subtraction of the $\om$-pole,
namely
\begin{equation}
  \label{eq:reg_part}
  A_{gq}(\om) \equiv \Ga_{gq,0}(\om) - \frac{2\CF}{\om} \;, \qquad
  A_{gg}(\om) \equiv \Ga_{gg,0}(\om) - \frac{2\CA}{\om} \;.
\end{equation}
Their values at $\om = 0$ are
\begin{equation}
  \label{eq:Azero}
  \begin{pmatrix}
    \Ga_{qq,0}(0) & \Ga_{qg,0}(0) \\[2ex] A_{gq}(0) & A_{gg}(0)
  \end{pmatrix}
  =
  \begin{pmatrix}
    0 & \frac{4 \Tf}{3} \\[2ex] -\frac{3 \CF}{2} & -\frac{11\CA + 4\Tf}{6}
  \end{pmatrix}
  \;.
\end{equation}

\section{Kernels and characteristic functions\label{a:kcf}}

Our method of resumming energy-scale dependent terms relies on the introduction
of improved kernels whose characteristic functions%
\footnote{Apart from the (running) coupling factors, we always deal with
  scale-invariant kernels.}
$\chi_{\om}(\ga)$ are $\om$-dependent.  In general such characteristic functions
are symmetric in the $\ga \to 1+\om-\ga$ transformation and have the following
structure:
\begin{equation}\label{eq:chi_struct}
 \frac1{\om} \chi_{\om}(\ga) =
 \mathcal{M}(\om) \left[ \chi_L(\ga) + \chi_L(1+\om-\ga) \right] \;,
\end{equation}
where the left-projection $\chi_L$ contains collinear (and possibly
higher-twist) singularities only in the half-plane $\Re(\ga) \leq 0$.%
\footnote{Note that in this article we adopt the asymmetric --- {\em upper} in
  the notations of~\cite{rgiggf} --- energy scale $s_0 = k^2$, since $k^2/s$ is
  the correct Bjorken scaling variable in the collinear limit $k \gg k'$ we are
  interested to. This causes the $\om$-shift to apply only on the $(1-\ga)$
  argument of $\chi_L$ and asymmetric kinematical constraints in the $z$
  variable as shown below.}

The $\om$-dependence in the argument of the second $\chi_L$ term imposes
kinematical constraints on the longitudinal momentum fraction variable $z$
conjugated to $\om$. In fact, by denoting by $zM(z)$ and $K_L$ the inverse
Mellin transforms of $\mathcal{M}$ and $\chi_L$ respectively, we have
\begin{equation}\label{eq:K_z_kt}
 K(z;\kt,\kt') \equiv \frac1{k^2} \int\frac{\dif\om}{2\pi\ui} \; z^{-\om}
 \int\frac{\dif\ga}{2\pi\ui} \; \left(\frac{k^2}{k'{}^2}\right)^{\ga} \;
 \frac1{\om}\chi_{\om}(\ga) = z' M(z') K_L(k_>,k_<) \;,
\end{equation}
where $k_< \equiv \min(k,k')$, $k_> \equiv \max(k,k')$ and
$z' \equiv z \cdot \max(1,k'{}^2/k^2)$. Since $0<z'<1$, the kinematical constraint
$k'{}^2 < k^2/z$ follows.

The lowest-order matrix kernel $\cK_0$ in $(z,\kt)$-space can be derived from
eq.~(\ref{eq:chi_mat_0_all_prerequisites}) and reads
\begin{align}\label{K0zk}
 [\alh\cK_0](z;k,k') &= \alh(q^2)
 \begin{pmatrix}
   0 & 0 \\[2ex] 0 & 2\CA K_0(z;k,k')
 \end{pmatrix}
 \\ \nonumber
 &+ \alh(k_>^2) \left[ z'
 \begin{pmatrix}
   P_{qq,0}(z') & P_{qg,0}(z') \\[2ex] P_{gq,0}(z') & P_{gg,0}(z') -\frac{2\CA}{z'}
 \end{pmatrix}
 K_c(k,k') + z'
 \begin{pmatrix}
   0 & \Dqg(z') \Kht(k,k') \\[2ex] 0 & 0
 \end{pmatrix}
 \right]
\end{align}
having defined
\begin{align}
 K_c(k,k')  &\equiv \frac1{k_>^2}\,,
 \label{K_c_bis} \\
 \Kht(k,k') &\equiv \frac{2}{3} \frac{k_<^2}{k_>^4}\,,
 \label{K_ht} \\
 \Dqg(z) &\equiv \dqg \, 3 (1-z)^2 \;.
 \label{Dqgz} 
\end{align}
The terms in square brackets in eq.~(\ref{K0zk}) correspond to the operator
$\cK_{\mathrm{coll}}$ introduced in eq.~(\ref{integralEq}). The action of the
first term $\sim\alh(q^2)$ on a test function $f(x,k)$ is
\begin{align}\label{K0action}
 \int_x^1 \frac{\dif z}{z}\int\dif k'{}^2 \;&\alh(q^2) K_0(z;k,k')
 f\big(\frac{x}{z},k'\big)
 \\ \nonumber
 &\equiv \int_x^1 \frac{\dif z}{z}\int
 \frac{\dif^2 \qt}{\pi q^2} \;\alh(q^2) \left[ f\big(\frac{x}{z},|\kt+\qt|\big)
 \Theta(k^2-z k'{}^2) - \Theta(k-q)f\big(\frac{x}{z},k\big) \right] \;,
\end{align}
while the action of the terms $\sim\alh(k_>^2)$, e.g.\ the higher-twist one, is
\begin{align}\label{Khtaction}
 \int_x^1 &\frac{\dif z}{z}\int\dif k'{}^2 \;\alh(k_>^2) z'\Dqg(z') \Kht(k,k')
 f\big(\frac{x}{z},k'\big)
 \\ \nonumber
 \equiv
 &\int_x^1 \frac{\dif z}{z} \bigg\{ \int_0^{k^2} \dif k'{}^2 \;\alh(k^2) z\Dqg(z)
  + \int_{k^2}^{k^2/z} \dif k'{}^2 \; \alh(k'{}^2) z\frac{k'{}^2}{k^2}
   \Dqg\big(z\frac{k'{}^2}{k^2}\big) \bigg\} \Kht(k,k') f\big(\frac{x}{z},k') \;.
\end{align}

In order to obtain $\cK_1(z;k,k')$ according to
eq.~(\ref{eq:K1ansatz}), we start by
computing the first term proportional to $\chi_c \to K_c$.  The inverse Mellin
transform of $\Ga_1^{(\MSbar)}$ is just the matrix of the two-loop singlet
splitting functions in the $\MSbar$-scheme~\cite{NLOsplit}.  The inverse Mellin
tranform of the subtraction $\cK_0^{(1)} \cK_0^{(0)}$ can be either computed by
inverting the expressions listed in eq.~(\ref{K01K00ij}), or by convolution in
$z$-space of the corresponding factors. Here we choose the second method, by
computing first the analytic expressions of all factors in eq.~(\ref{K01K00ij})
and then the numerical convolution of the ensuing functions. We already obtained
the Mellin transform of $\Dqg$ in eq.~(\ref{Dqgz}); the
transforms of the $\Ga_{ij}$ factors are just the one-loop splitting
functions reported in eqs.~(\ref{gamma0_qq}-\ref{gamma0_gg}); the remaining
functions are listed below:
\begin{align}
 &\int\frac{\dif\om}{2\pi\ui} \; z^{-\om} \; c_c(\om) = z\,,
 \label{inv_cc} \\
 &\int\frac{\dif\om}{2\pi\ui} \; z^{-\om} \; \frac{2\CA}{\om} c_0(\om)
  = 2\CA \log(1-z)\,,
 \label{inv_c0} \\
 &\int\frac{\dif\om}{2\pi\ui} \; z^{-\om} \; \chiht(0,\om)
  = \frac{2}{3}\left [\delta(1-z) + z^2 \right]\,.
 \label{inv_chiht}
\end{align}

Finally, we need the inverse Mellin transform
\begin{equation}\label{inv_Dp}
 \int\frac{\dif\om}{2\pi\ui} \; z^{-\om} \;
 \left(\frac1{\om}-\frac2{1+\om}\right) = 1 - 2 z\,,
\end{equation}
and the kernel
\begin{equation}
 K_{1,\mathrm{reg}} \equiv \tilde{K}_1 - \tilde{\chi}_1^{(0)} K_c,
\end{equation} 
 whose $\om$-shifted form occurs directly in eq.~(\ref{runningKernel}), and
 has characteristic function 
\begin{align}\label{eq:chireg}
 \chireg &\equiv \tilde{\chi}_1 - \tilde{\chi}_1^{(0)} \chi_c
 \\ \nonumber
 &= \chi_1 - \chi_1^{\run}
  - \chi_0 \left( \dot{\chi}_0 + \chi_c \frac{A_{gg}(0)}{2\CA} \right)
  - \frac{\CF}{\CA} \chi_c \left( \chi_c \frac{\Ga_{qg,0}(0)}{2\CA}
    + \chiht \frac{\dqg}{2\CA} \right) -\tilde{\chi}_1^{(0)}\chi_c \;.
\end{align}

The numerical coefficients $A_{gg}(0), \Ga_{qg,0}(0)$ can be found in
eq.~(\ref{eq:Azero}), $\dqg \equiv \Delta_{qg}(0)$ in eq.~(\ref{eq:Delta_qg_MSbar}) and
$\tilde{\chi}_1^{(0)}$ in eq.~(\ref{chi1t0}). Furthermore, the
computation of the kernel $K_{1,\mathrm{reg}}$ requires the subtraction from the
NL$x$ BFKL kernel~\cite{FaLi98,CaCi98} of the running coupling terms and of additional
kernels corresponding to the characteristic functions on the r.h.s. of
eq.~(\ref{eq:chireg}). They are given by
\begin{align}
 \chi_c &\to \frac1{k_>^2}\,,
 \label{K_c} \\
 \chi_0 \dot{\chi}_0 &\to -\frac1{4|k^2-k'{}^2|}
 \left[\log^2 \frac{k'{}^2}{k^2} + 4\Li\Big(1-\frac{k_<^2}{k_>^2}\Big)\right]\,,
 \label{K_0dot0} \\
 \chi_0\chi_c &\to \frac1{k_>^2} \log \Big(\frac{k_>^2}{k_<^2}-1\Big)
 - \frac1{k_<^2} \log \Big(1-\frac{k_<^2}{k_>^2}\Big)\,,
 \label{K_0c} \\
 \chi_c^2 &\to \frac1{k_>^2}\left( \log\frac{k_>^2}{k_<^2} + 2 \right)\,,
 \label{K_c_sq} \\
 \chi_c \chiht &\to \frac1{k_>^2}\left( 1
  - \frac1{3}\frac{k_<^2}{k_>^2} \right)\,.
 \label{K_c_ht}
\end{align}
The resulting expression for $K_{1,\mathrm{reg}}$ is
\begin{align}
  K_{1,\mathrm{reg}}(k,k') &= \frac14 \left\{
 \left(\frac{67}{9} - \frac{\pi^2}{3} -\frac{20 \Tr\nf}{9\CA} \right)
 \langle K_0 \rangle(k,k')
  + \frac{1}{{k'}^2 + k^2} \left[ \frac{\pi^2}{3} +
  4 \Li\big( \frac{k_{<}^2}{k_{>}^2}\big)\right]
  + \right. \nonumber \\
 & - \frac{1}{32} \left( 1 + \frac{2\Tr\nf}{\CA^3} \right)
  \left[\frac{2}{{k'}^2} + \frac{2}{k^2} +
  \left(\frac{1}{{k'}^2} - \frac{1}{k^2} \right)
  \log\left(\frac{k^2}{{k'}^2}\right)\right] + \nonumber \\
 & \left.
 - \left[3 + \left(\frac{3}{4} - \frac{({k'}^2+k^2)^2}{32{k'}^2 k^2}\right)
  \right]  \left( 1 + \frac{2\Tr\nf}{\CA^3} \right)
  \int_0^{\infty} \frac{\dif y}{k^2 + y^2 {k'}^2} \;
  \log\left|\frac{1+y}{1-y}\right|
  \right\} + \nonumber \\
 & + \frac{3}{2} \zeta(3) \delta(k^2-k'^2)
  + \frac{4 \Li(1-k_{<}^2/k_{>}^2)}{|{k'}^2 - k^2|} + \nonumber \\
 & -4 \frac{A_{gg}(0)}{2\CA}{\rm sgn}({k}^2-{k'}^2)
  \left( \frac{1}{k^2} \log\frac{|{k'}^2-k^2|}{{k'}^2} -
  \frac{1}{{k'}^2} \log\frac{|{k'}^2-k^2|}{{k}^2} \right) + \nonumber \\
 & -\frac{\CF}{\CA} \frac1{k_>^2} \left[
    \frac{\Ga_{qg,0}(0)}{2\CA}
    \left( \log\frac{k_>^2}{k_<^2} + 2 \right)
    + \frac{\dqg}{2\CA}\left( 1 - \frac1{3}\frac{k_<^2}{k_>^2} \right)
  \right]
    - \tilde{\chi}_1^{(0)} \frac1{k_>^2} \;.
 \label{K1reg}
\end{align}
where $\langle K_0 \rangle$ denotes the azimuthal average of the LL$x$ BFKL
kernel whose action on a test function $f(k)$ is given by
\begin{equation}\label{azimK0}
 [\langle K_0 \rangle f](k) = \int\dif k'{}^2 \; \frac{1}{|{k'}^2-k^2|}
 \left [f(k') - \frac{2 k_{<}^2}{k'{}^2 + k^2} f(k) \right] \;.
\end{equation}

Finally, we provide the eigenvalue function of $K_{1,\mathrm{reg}}$ with
kinematical constraints, which is given by
$\chireg{}_{,L}(\ga) + \chireg{}_{,L}(1+\om-\ga)$, and occurs directly in
eq.~(\ref{runningKernel}).   The left projection
$\chireg{}_{,L}$ of the eigenvalue function (\ref{eq:chireg}) can be computed
starting from the expression of $\tilde{\chi}_{1L}$ in eq.~(A.13) of
ref.~\cite{rgiggf} and noticing that:\\
{\it(i)} here we have more terms to subtract, namely those proportional to
$\CF/\CA$ and $\tilde{\chi}_1^{(0)}$;\\
{\it(ii)} in ref.~\cite{rgiggf} we subtracted the $\nf$-part of the double pole
by letting $A_{gg}(0) \to 2\CA A_1(0) \equiv A_{gg}(0) + (\CF/\CA) \Ga_{qg,0}(0)$,
while here we keep only $A_{gg}(0)$ in front of $\chi_0\chi_c$ since the
$\nf$-dependent double pole is subtracted by the $\Ga_{qg}\chi_c^2$ term.

Therefore, in order to complete the calculation we need the left projections of
the following kernels:
\begin{align}
 [\chi_c^2]_L(\ga) &= [\chi_{cL}(\ga)]^2 + 2 \chi_{cL}(\ga)
 =  \frac1{\ga^2} + \frac{2}{\ga}\,,
 \\
 [\chi_c \chiht]_L(\ga) &= \chi_{cL}(\ga) - \frac1{2}\chiht{}_L(\ga)
 = \frac1{\ga} - \frac1{3(1+\ga)} \;.
\end{align}
The final result is
\begin{align}
 \chireg{}_{,L}(\ga) &= [\psi(1)-\psi(\ga)]
 \left[\psi'(\ga) - \frac{A_{gg}(0)}{2\CA}\chi_c(\ga)
 + \frac{67}{36}-\frac{\pi^2}{12}-\frac{5}{18}\frac{\nf}{\CA} \right]
 \nonumber \\
 &\quad + \frac12\psi''(\ga) + \Pi(\ga) - \Phi_{L}(\ga)
 +\frac{\pi^2}{8} \left[\psi\Big(\frac{1+\ga}{2}\Big)
 -\psi\Big(\frac{\ga}{2}\Big) \right] + \frac{3}{4}\zeta(3)
 \nonumber \\
 &\quad +\frac1{32} 
  \left\{
    -3M(\ga) + \left(1+\frac{\nf}{\CA^3}\right)
      \left[ \frac14\left(\frac1{\ga^2}-\frac1{(1-\ga)^2}\right)
             -\frac12\left(\frac1{\ga}-\frac1{1-\ga}\right)
      \right.
  \right.
 \nonumber \\
 &\qquad\qquad\qquad\qquad\qquad\qquad\qquad\;
 \left.\left. 
             +\frac1{32}\Big(M(\ga+1)+M(\ga-1)\Big)
             -\frac{11}{16}M(\ga)
       \right]
 \right\}
 \nonumber \\
 &\quad -\frac{\CF}{2\CA^2}\left[ \Ga_{qg,0}(0)
  \left( \frac1{\ga^2} + \frac{2}{\ga} \right)
 + \dqg \left( \frac1{\ga} - \frac1{3(1+\ga)} \right)
 \right] - \frac{\tilde{\chi}_1^{(0)}}{\ga} \;,
 \label{chi1regL}
\end{align}
where
\begin{align}
\label{d:Pi}
 \Pi(\ga) &\equiv \int_0^1 dt \; t^{\ga-1}
 \frac{\Li(1)-\Li(t)}{1-t} = \sum_{n=0}^\infty \frac{\psi'(n+1)}{n+\ga}\,, \\
\label{d:PhiL}
 \Phi_{L}(\ga) &\equiv \sum_{n=0}^\infty (-1)^n
 \frac{\psi(n+1+\ga)-\psi(1)}{(n+\ga)^2}\,, \\
\label{d:fc}
 M(\ga) &\equiv \frac1{\ga-\half} \left[
 \psi'\Big(\frac{1+\ga}{2}\Big) - \psi'\Big(\frac{\ga}{2}\Big)
 + \psi'\Big(\frac{1}{4}\Big) -\psi'\Big(\frac{3}{4}\Big) \right] \;.
\end{align}
The explicit form of $K_{1,\mathrm{reg}}$ in ($\kt, z$) space is obtained by
use of eq.~(\ref{eq:K_z_kt}), or by introducing the kinematical constraints on
eq.~(\ref{K1reg}) directly.  The final form of $\cK_1(z; \kt, \kt')$ follows
from eq.~(\ref{runningKernel}).


\end{document}